\documentclass[aps,amssymb,amsmath,pra,twocolumn,showpacs,superscriptaddress]{revtex4-1}

\usepackage{graphicx}
\usepackage{dcolumn}
\usepackage{bm}
\usepackage{subfigure}
\usepackage{color}

\usepackage{amssymb,amsmath,amsfonts,latexsym,graphicx,verbatim}

\usepackage[margin=0.6in]{geometry}

\usepackage[english]{babel}
\usepackage{times}
\usepackage{latexsym}
\usepackage{fancyhdr}
\usepackage{float}
\usepackage{afterpage}

\usepackage{listings}
\usepackage{multirow}

\usepackage{bbm}
\usepackage{upgreek}

\definecolor{Ablue}{rgb}{0.96,0.24,0.00}

\definecolor{Abluetitle}{rgb}{0.,0.24,0.51}
\newcommand{\bluetitle}{\color{Abluetitle}}

\definecolor{orange}{rgb}{0.96,0.24,0.00}

\definecolor{darkred}{rgb}{0.55, 0.0, 0.0}

\definecolor{Gray}{gray}{0.85}
\definecolor{LightCyan}{rgb}{0.88,1,1}
\definecolor{darksalmon}{rgb}{0.91, 0.59, 0.48}
\definecolor{maroon}{cmyk}{0,0.87,0.68,0.32}

\definecolor{mustard}{rgb}{1.0, 0.86, 0.35}
\newcolumntype{a}{>{\columncolor{Gray}}c}
\newcolumntype{b}{>{\columncolor{white}}c}

\usepackage{array}
\newcolumntype{L}[1]{>{\raggedright\let\newline\\\arraybackslash\hspace{0pt}}m{#1}}
\newcolumntype{C}[1]{>{\centering\let\newline\\\arraybackslash\hspace{0pt}}m{#1}}
\newcolumntype{R}[1]{>{\raggedleft\let\newline\\\arraybackslash\hspace{0pt}}m{#1}}

\usepackage[colorlinks=true , citecolor=blue,urlcolor=blue]{hyperref}

\newcommand{\xa}{\alpha}

\newcommand{\xb}{\beta}

\newcommand{\xg}{\gamma}
\newcommand{\xt}{\theta}

\newcommand{\xo}{\omega}
\newcommand{\xph}{\phi}

\newcommand{\app}{\approx}
\newcommand{\dw}{\downarrow}
\newcommand{\up}{\uparrow}
\newcommand{\trepol}{t_{\R{repol}}}

\newcommand{\Cs}{{}^{13}\R{C}}

\newcommand{\mB}[0]{\mathcal{B}}
\newcommand{\dxo}[0]{\dot\omega}

\newcommand{\xD}{\Delta}

\newcommand{\xO}{\Omega}

\newcommand{\fr}[2]{\frac{#1}{#2}}

\newcommand{\wt}[1]{\widetilde{#1}}

\newcommand{\sq}[1]{\sqrt{#1}}

\newcommand{\mH}[0]{\mathcal{H}}

\newcommand{\rt}{\rightarrow}

\newcommand{\beq}{\begin{equation}}
\newcommand{\eeq}{\end{equation}}
                  
\newcommand{\benum}{\begin{enumerate}}
\newcommand{\eenum}{\end{enumerate}}
                    
\newcommand{\bit}{\begin{itemize}}
\newcommand{\eit}{\end{itemize}}

\newcommand{\bea}{\begin{eqnarray}}
\newcommand{\eea}{\end{eqnarray}}

\newcommand{\bev}{\begin{verbatim}}
\newcommand{\eev}{\end{verbatim}}

\newcommand{\lcb}{\left\{}
\newcommand{\rcb}{\right\}}

\newcommand{\T}[1]{\textbf{#1}}
\newcommand{\I}[1]{\textit{#1}}
\newcommand{\R}[1]{\textrm{#1}}

\newcommand{\zfl}[1]{\protect\label{fig:#1}}
\newcommand{\zfr}[1]{Fig. \ref{fig:#1}}

\newcommand{\zsl}[1]{\label{sec:#1}}

\newcommand{\ket}[1]{\left\vert{#1}\right\rangle}

\newcommand{\ba}{\left\{ \begin{array}{lr}}
\newcommand{\ea}{\end{array}\right.}

\newcommand{\blist}[1]{
 \begin{list}{#1}
 \begin{align}
	 arrow
 \end{align}
 $\checkmark\star
  { \setlength{\itemsep}{3pt}
     \setlength{\parsep}{2pt}
     \setlength{\topsep}{3pt}
     \setlength{\partopsep}{0pt}
     \setlength{\leftmargin}{1em}
     \setlength{\labelwidth}{1em}
     \setlength{\labelsep}{0.5em} } }
\newcommand{\elist}{
  \end{list}  }

\DeclareMathSymbol{\vartheta}{\mathalpha}{letters}{"12}
\DeclareMathSymbol{\theta}{\mathalpha}{letters}{"23}
\DeclareMathSymbol{\phi}{\mathalpha}{letters}{"27}
\DeclareMathSymbol{\varphi}{\mathalpha}{letters}{"1E}

\newcommand{\bef}
{
\begin{figure}[htbp]
\centering
}

\newcommand{\eef}{\end{figure}}

\newcommand{\beginsupplement}{%
        \setcounter{table}{0}
        \renewcommand{\thetable}{S\arabic{table}}%
        \setcounter{figure}{0}
        \renewcommand{\thefigure}{S\arabic{figure}}%
     }

\newcommand{\affA}{Department of Chemistry, University of California Berkeley, and Materials Science Division Lawrence Berkeley National Laboratory, Berkeley, California 94720, USA.}
\newcommand{\affB}{Department of Physics, CUNY-City College of New York, New York, NY 10031, USA.}
\newcommand{\affC}{CUNY-Graduate Center, New York, NY 10016, USA.}
\newcommand{\affD}{Department of Chemical and Biomolecular Engineering, and Materials Science Division Lawrence Berkeley National Laboratory University of California, Berkeley, California 94720, USA.}
\newcommand{\affE}{Fakult\"{a}t Physik, Technische Universit\"{a}t Dortmund, D-44221 Dortmund, Germany.}
\newcommand{\affF}{Department of Physics, University of California Berkeley, Berkeley, California 94720, USA.}

\newcommand{\affH}{Department of Physics and Astronomy, Dartmouth College, Hanover, New Hampshire 03755, USA.}

\begin{document}
\title{\bluetitle{Enhanced dynamic nuclear polarization via swept microwave frequency combs}}
\author{A. Ajoy}\email{ashokaj@berkeley.edu}\affiliation{\affA}
\author{R. Nazaryan}\affiliation{\affA}
\author{K. Liu}\affiliation{\affA}
\author{X. Lv}\affiliation{\affA}
\author{B. Safvati}\affiliation{\affF}
\author{G. Wang}\affiliation{\affA}
\author{\\E. Druga}\affiliation{\affA}
\author{J. A. Reimer}\affiliation{\affD}
\author{D. Suter}\affiliation{\affE}
\author{C. Ramanathan}\affiliation{\affH}
\author{C. A. Meriles}\affiliation{\affB}\affiliation{\affC}
\author{A. Pines}\affiliation{\affA}


\begin{abstract}
 Dynamic Nuclear Polarization (DNP) has enabled enormous gains in magnetic resonance signals and led to vastly accelerated NMR/MRI imaging and spectroscopy. Unlike conventional \I{cw}-techniques, DNP methods that exploit the full electron spectrum are appealing since they allow direct participation of all electrons in the hyperpolarization process. Such methods typically entail sweeps of microwave radiation over the broad electron linewidth to excite DNP, but are often inefficient because the sweeps, constrained by adiabaticity requirements, are slow. In this paper we develop a technique to overcome the DNP bottlenecks set by the slow sweeps, employing a swept microwave frequency comb that increases the effective number of polarization transfer events while respecting adiabaticity constraints. This allows a multiplicative gain in DNP enhancement, scaling with the number of comb frequencies and limited only by the hyperfine-mediated electron linewidth. We demonstrate the technique for the optical hyperpolarization of $\Cs$ nuclei in powdered microdiamonds at low fields, increasing the DNP enhancement from 30 to 100 measured with respect to the thermal signal at 7T. For low concentrations of broad linewidth electron radicals, e.g. TEMPO, these multiplicative gains could exceed an order of magnitude. 
\end{abstract}

\maketitle

\T{\I{Introduction:}} -- Dynamic nuclear polarization (DNP) -- the process of polarizing (\I{cooling}) nuclear spins to a spin temperature far lower than the lattice temperature~\cite{Carver53,Abragam78} -- has emerged as a technological breakthrough that serves as the starting point for a wide-range of applications, including signal enhanced spectroscopy~\cite{Wind85,Maly08} and imaging~\cite{Mccarney07} and for state initialization in quantum information processing and metrology~\cite{Kane00,Foletti09}. Indeed, magnetic resonance (NMR and MRI) signals from \I{hyperpolarized} nuclear spins can be enhanced by several orders of magnitude allowing enormous gains, even approaching a million-fold, in experimental averaging time. This has opened up avenues for the sensitive probing of phenomena, species and surfaces~\cite{Lesage10}, whose detection would otherwise have remained intractable.

\begin{figure}[t]
  \centering
  \includegraphics[width=0.49\textwidth]{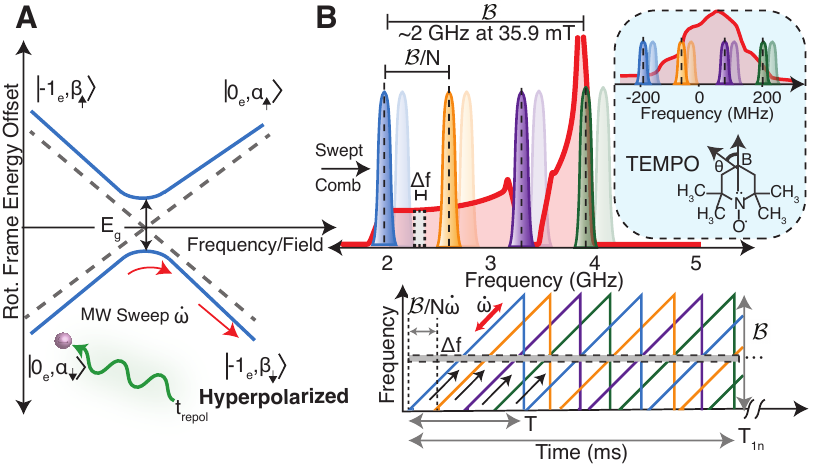}
 \caption{\textbf{Frequency comb enhanced DNP.} (A) Hyperpolarization processes via frequency/field swept techniques are effectively Landau-Zener traversals of level anti-crossings in a dressed electron-nuclear basis. The electron is repolarized every $\trepol$, and sweeps with an adiabatic scan rate $\dxo$ lead to polarization transfer. $E_g$ refers to the energy gap. Levels shown here are for the NV center - $\Cs$ spin system that we shall consider in detail in this paper. (B) \I{Principle.} Microwave frequency comb sweeping the entire inhomogeneously broadened electron spectrum (with linewidth $\mB$)  allows a repeated polarization transfer event with every successive comb frequency, and produces a {multiplicative} boost in DNP enhancement. Red shaded area shows the spectrum for NV center electrons in diamond powder at 35.9mT.   The $N$ comb frequencies can be as close as $\xD f$, the hyperfine mediated linewidth (shaded), and sweep every electron packet as often as $\trepol$. \I{Inset:} similar method could be applied to broad line electron radicals like TEMPO, shown here 3.35T with the spectrum centered at 95GHz~\cite{Feintuch11}. \I{Lower panel:} time domain implementation through multiple cascaded frequency sweepers, illustrating the ability to maintain the adiabatic rate $\dxo$ while increasing the effective number of sweeps by $N$.}
\zfl{lac}
\end{figure}

In its simplest manifestation DNP involves the use of electrons whose polarization is transferred to the nuclear spins via microwave irradiation~\cite{Goldman}, allowing a polarization enhancement $\varepsilon \lesssim \xg_e/\xg_n$, where $\xg_{e,n}$ are the gyromagnetic ratios of the electron and nuclear spins respectively. Resonant polarization transfer between electron and nuclear spin is achieved via microwave excitation.  Depending on the concentrations of the electron and nuclear spins in the insulating solid, the transfer can be mediated by thermal mixing, the cross effect, the solid effect and even the Overhauser effect. However several common (e.g. nitroxide based) electron polarizing agents have large \I{g}-anisotropy and severely inhomogeneously \I{broadened} electronic linewidths that scale rapidly with field and can be as broad as 0.5GHz at high fields ($>$3T) ~\cite{Feintuch11,Song06,Hu07,Armstrong07}. This broadening limits the number of spins contributing to the resonant energy exchange at a particular microwave frequency. Similar problems can exist even at low fields for some systems. For instance, Nitrogen Vacancy (NV) center defects in diamond~\cite{Wrachtrup06,Doherty12} have garnered much attention as optical hyperpolarizing agents because the NV electrons can be \I{fully} optically polarized at room temperature~\cite{Fischer13}, opening the possibility for DNP enhancements larger than traditional bounds set by the gyromagnetic ratios, without the need for cyrogens. Interest has been particularly focused on ``\I{hyperpolarized nanodiamonds}", because their inherently high surface area makes them attractive for the optical hyperpolarization of liquids brought in contact with them~\cite{chen15b, Andrich14}. While this has been a long-sought goal, technical challenges presented by the NV electrons make the production of hyperpolarized nanodiamonds challenging. In particular, the spin-1 NV centers have significant broadening on account of different crystallite orientations having different frequencies, giving rise to spectra broadened by $>$1GHz even at modest (30mT) fields. Unsurprisingly, precise energy matching to the nuclei in all these situations is challenging to achieve. Indeed DNP traditionally has relied largely on \I{cw}-microwave techniques (solid and cross-effects) where a single frequency is saturated~\cite{Hovav10,Shimon12}, and consequently for static samples only a small fraction of the broad electron spectrum directly contributes to the obtained enhancement.

In principle however, significant gains in polarization enhancements can be achieved by exploiting the \I{full} broad electron linewidth for DNP via more sophisticated quantum control on the electron spins, wherein every electron ``packet'' directly  contributes to the DNP process~\cite{Hovav14}. In this paper we shall demonstrate a strategy to achieve this for the case of strongly anisotropic radicals in the limit of low ({dilute}) concentrations, where inter-electron couplings can be neglected -- a situation pertinent for a wide class of nitroxide radicals, and endogenously radicals native to several systems~\cite{Guy17,Rej17}. Since savings in experimental time scale $\propto \varepsilon^2$, methods to increase hyperpolarization efficiency will directly translate to accelerated spectroscopy and imaging. Indeed, a surge in recent interest in DNP control techniques has been fueled by advances in instrumentation (sources~\cite{Guy15,Yoon16} and synthesizers~\cite{Kaminker17}) that enable the rapid and coherent manipulation of electrons at high fields~\cite{Hoff15,Can15,Can17}. Particularly attractive amongst them is the use of frequency or field swept techniques, e.g. integrated solid effect (ISE)~\cite{Henstra88b, Henstra90, Tateishi14}, that are suited to exploiting the wide electron bandwidth while only requiring modest microwave power.

\T{\I{Principle:}} -- The DNP process underlying these techniques can be simplistically described as traversals of a level anti-crossing (LAC) in an electron-nuclear dressed basis (\zfr{lac}). Polarization transfer occurs via Landau-Zener (LZ) tunneling~\cite{Zener32}, the onus of thermal contact being placed on maintaining adiabaticity during the sweep~\cite{Rubbmark81}. The DNP transfer efficiency, governed by the tunneling probability, is given by $\varepsilon\propto \exp(-E_g^2/\dot\xo)$, where $\dot\xo$ is the sweep rate and the $E_g$ is the effective energy gap, and depends on several parameters including the electron Rabi frequency $\xO_e$~\cite{Henstra88b}, hyperfine coupling to the target nucleus, and orientation. Despite harnessing the full electron linewidth, the frequency sweeps are often slow, and the requirement of adiabaticity sets bounds on the rate of polarization transfer. To illuminate this in more detail, let us assume an inhomogeneous electron linewidth ${\cal B}$, leading to a single traversal time $T=\mB/\dot\xo$. Each electron frequency packet, however, has repolarized within a time $t_{\R{repol}}\leq T_{1e}\ll T$, and is available again for DNP transfer,  but instead has to wait the full period $T$ when the subsequent sweep leads to the next polarization transfer event. Since the nuclear polarization is proportional to the total number of sweeps $T_{1n}/T$, the slow sweeps set a \I{bottleneck} on the DNP process, since an increasing bandwidth $\mB$ leads to a longer period $T$. For instance, for the typical case of TEMPO at 3.35T and 50K, $\mB\app$ 0.5GHz and considering $\xO_e$=1MHz~\cite{Feintuch11,Siaw16}, $T$=500ms, which far exceeds the inherent repolarization time, $T\gtrsim \mB/\xO_e^2\gg T_{1e}\app$ 1ms~\cite{Feintuch11,Shimon12}. 


\begin{figure}[t]
  \centering
  \includegraphics[width=0.5\textwidth]{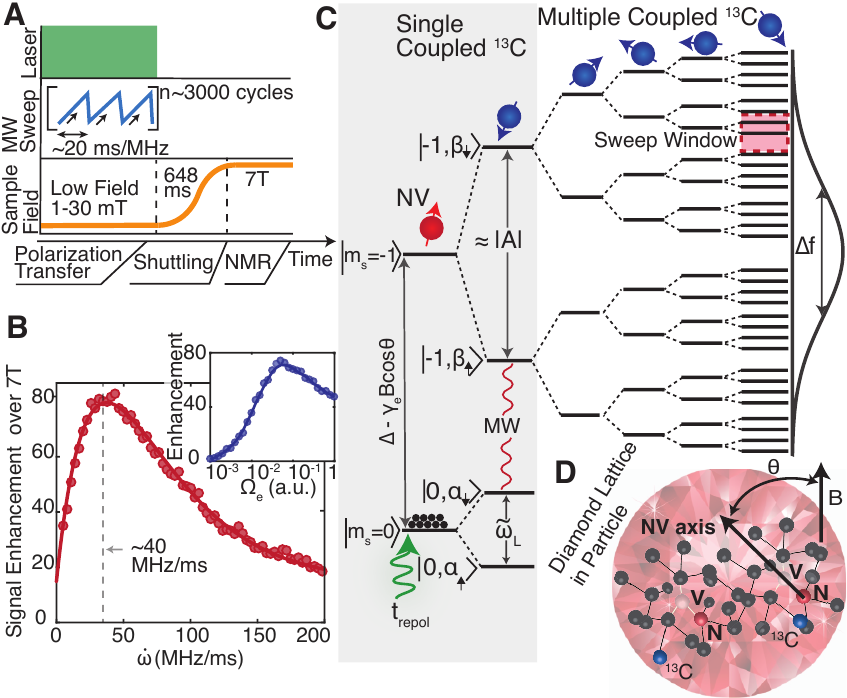}
  \caption{\textbf{$\Cs$ Hyperpolarization in diamond powder.} (A) \I{Sequence of events.} Room temperature $\Cs$ DNP from optically pumped NV centers is achieved via microwave sweeps at low field $B$=1-30mT under continuous 532nm optical illumination. Bulk polarization is inductively detected by sample shuttling to 7T. (B) \I{Sweep rate.} DNP dependence on MW sweep rate shows an optimal $\dxo$ set by adiabaticity constraints. \I{Inset:} Dependence on Rabi frequency $\xO_e$. (C) \I{DNP Mechanism.} Energy levels of a NV electron and a single $\Cs$ nuclear spin hyperfine-coupled with $A$ (grey box), in the low field regime where nuclear Larmor frequency $\xo_L\lesssim |A|$. For simplicity the $m_s=+1$ manifold is not shown. Swept microwaves lead to sequential excitation of LZ crossings and consequent $\Cs$ hyperpolarization. For an NV center coupled to multiple $\Cs$ nuclei one obtains a broadened ESR line by $\xD f$. Sweeping over any window (shaded) leads to hyperpolarization with the signal proportional to the local density of states.  (D) \I{Diamond lattice in a microparticle.} NV axes are randomly oriented with respect to the field $B$.}
\zfl{theory}
\end{figure}

\begin{figure}[t]
  \centering
  \includegraphics[width=0.49\textwidth]{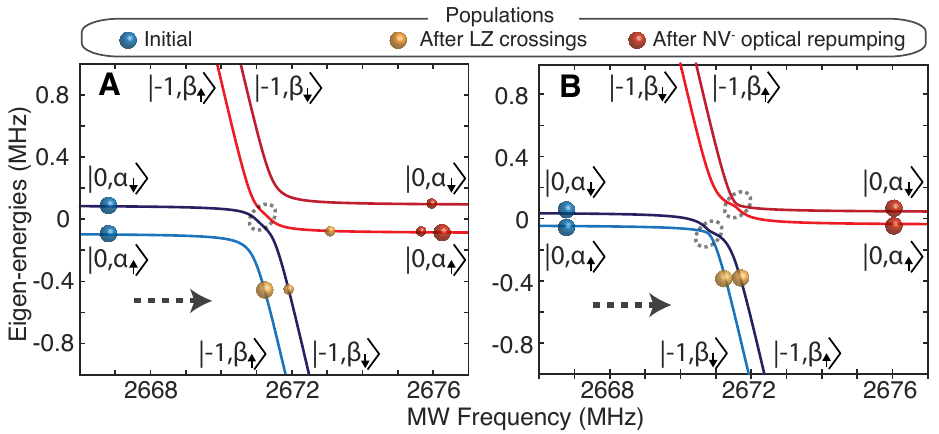}
  \caption{\textbf{Mechanism of polarization transfer.} Calculated energy diagram in the rotating frame for an NV electron spin hyperfine-coupled to a single $\Cs$ nuclear spin (grey box in \zfr{theory}C), and corresponding to the $m_S=0\leftrightarrow m_S=-1$ subset of transitions~\cite{Ajoy17, Zangara18}. (A) and (B) panels assume a hyperfine coupling $A_{zz}=+0.5$MHz and $-0.5$MHz respectively,  and $B$=10 mT, $\theta$=45 deg., and use a transverse hyperfine constant $A_{zx}=0.3|A_{zz}|$. Colored solid circles denote populations at different stages during a sweep in the direction of the arrow, and faint dashed circles indicate the narrower LACs where population transfer takes place. Panel illustrate that sweeping from low-to-high frequencies in the $m_s=-1$ manifold results in buildup of hyperpolarization in a direction aligned with the magnetic field for positive $A_{zz}$ (and analogous for $m_s=+1$ manifold and negative $A_{zz}$).}
\zfl{mechanism}
\end{figure}

In this paper, we demonstrate a simple method to overcome this bottleneck, increasing the effective number of polarization transfer events while maintaining the optimal adiabatic  sweep rates set by Landau-Zener conditions. Our method involves a swept microwave \I{frequency-comb}, that coherently and simultaneously sweeps the entire electron linewidth $\mB$ at $\dot\xo$, while maintaining adiabaticity for each sweep over an individual electron packet (see \zfr{lac}B). This allows repeated polarization transfer from each successive sweep of the comb, allowing one to gain a \I{multiplicative} DNP enhancement boost. Intuitively, the individual comb teeth can be as close as the electron linewidth $\xD f = 1/T_{2e}$ in frequency, and can sweep each electron packet as often as $t_{\R{repol}}$, allowing an enhancement gain $\varepsilon\rt N\varepsilon$. Since we work in the dilute electron limit, this electron packet linewidth predominantly arises due to hyperfine interactions with the surrounding nuclei. For the case of 10mM TEMPO for instance, this corresponds to a comb teeth separation of $1/T_M= $ 66kHz~\cite{Siaw14}, where $T_M$ is the phase memory time. Note that $T_M$ serves here as a lower bound for the electronic $T_{2e}$, since phenomena such as spectral and instantaneous diffusion reduce $T_M$ relative to $T_{2e}$~\cite{Schweiger01}. Without having to take into account specific details of the DNP mechanism in operation, one could bound the maximum enhancement gain $N\leq\R{min}\lcb \mB/\xD f,\mB/\dxo\trepol\app\mB/\xO_e^2T_{1e}\rcb$. The payoffs in hyperpolarization enhancements stemming from this multiplicative boost can be significant -- for TEMPO it could exceed an order of magnitude. More importantly, since the microwave power for each sweep remains $\xO_e$, the technique can be relatively easily implemented with existing technology -- the frequency comb being constructed by \I{time-cascading} sweeps from $N$ separate low-power amplifiers (\zfr{lac}B). 


While more generally employable, in this paper we demonstrate its application to $\Cs$ hyperpolarization in diamond particles via \I{optically} polarized electron spins associated with Nitrogen Vacancy (NV) center defects. We have recently developed a method for optical $\Cs$ DNP in powdered diamond at room temperature~\cite{Ajoy17}, employing a combination of laser and swept microwave irradiation at low magnetic fields ($B\sim$ 1-30mT). The DNP mechanism itself is a low-field complement to ISE, working in the regime where $\xo_L<|A|$, where $\xo_L=\xg_nB$ is the nuclear Larmor frequency. The NV centers are inhomogenously broadened to a powder pattern with bandwidth $\mB\app 2\xg_eB$, and here too the slow rate of microwave sweeps over $\mB$ limit the overall achievable nuclear polarization -- a challenge we overcome by the use of frequency combs. 

\begin{figure}[t]
  \centering
  \includegraphics[width=0.5\textwidth]{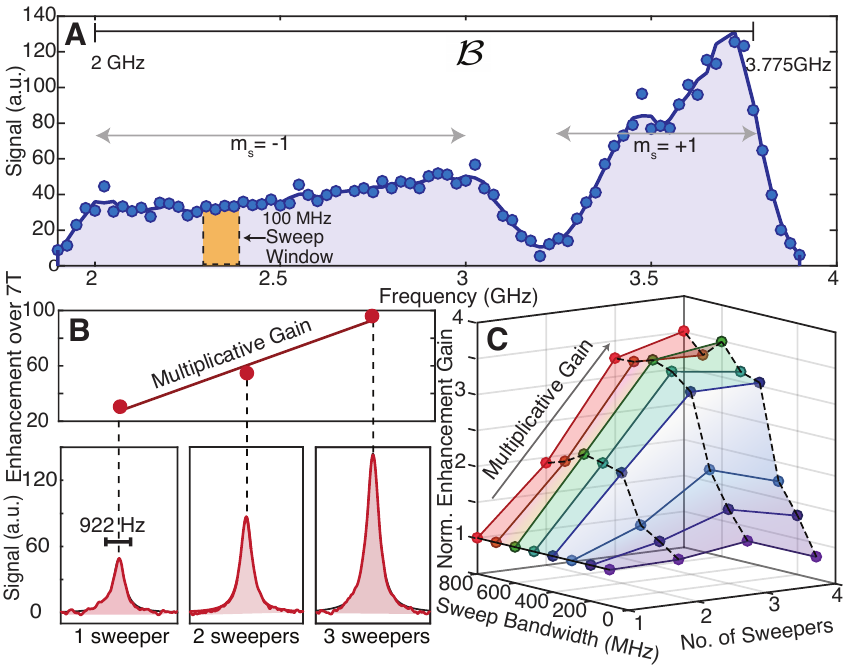}
  \caption{\textbf{Multiplicative DNP gains by frequency combs.} (A) \I{Indirect mapping of ESR lineshape.} Exemplary NV center powder pattern at $B$=27.7mT indirectly obtained via $\Cs$ DNP on a 100MHz window swept across the ESR line. Obtained signal is a convolution of the underlying electron spectrum with the sweep window. The ESR spectrum is orientationally broadened to $\mB\sim$1.8GHz. Sign of hyperpolarization is identical for the $m_s=\pm$1 manifolds and depends only on the direction of the MW sweep. (B) \I{Enhanced DNP gains.} Upper panel shows the multiplicative boost in the DNP enhancements employing a cascade of upto 3 microwave sweepers, over a 700MHz bandwidth at $B$=13mT. \I{Lower panel:} obtained hyperpolarized $\Cs$ spectra at 7T after 30 averages. (C) \I{Bandwidth dependence.} We characterize the multiplicative gain factor employing (upto) 4 cascaded sweepers, normalized by the use of a single one over the same band. Sweep bandwidths are centered at 2.8GHz in these experiments. }
\zfl{results}	
\end{figure}

\begin{figure}[t]
  \centering
	\includegraphics[width=0.5\textwidth]{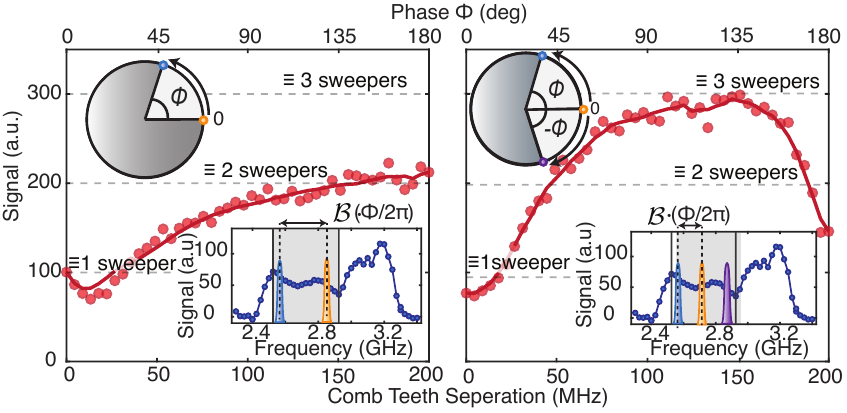}
        \caption{\textbf{Characterizing limits to multiplicative DNP gains} by determining the optimal frequency separation for an $N$ sweeper comb over a  $\mB$=400MHz bandwidth ($m_s$=-1 manifold) and fixed optimal sweep rate $\dxo=$39MHz/ms. \I{Inset:} shaded region denotes sweep bandwidth over the experimentally obtained powder pattern at 12mT. Sweepers are time cascaded with the ramps in \zfr{lac}B shifted by a phase $\xph$ (\I{upper axis}), translating to a frequency separation of the comb teeth by $f=(\xph/2\pi)\mB$ (\I{lower axis}). We study $\Cs$ DNP enhancements for (A) 2 and  (B) 3 sweepers, with phases set to $\{0,\xph\}$ and $\{-\xph,0,\xph\}$ respectively. \I{Insets} show them explicitly as phasors and in frequency domain. Data demonstrates that time ramps should be optimally phase shifted by $\xph_{\R{opt}}=2\pi/N$ for maximum DNP gains. As expected when $\xph=0$ there is no gain in using $N$ sweepers over a single one (lower dashed line), and for $\xph=180^{\circ}$ two and three sweepers provide similar enhancement (upper dashed line). Solid lines are guides to the eye.}
  \zfl{phase}	
\end{figure}

\T{\I{Low-field $\Cs$ DNP in diamond powder:}} -- To be more concrete, \zfr{theory}A presents the hyperpolarization sequence.  Laser irradiation polarizes the NV centers to the $m_s=0$ state, a feature that occurs independent of field. We estimate the resulting NV electron polarization to be close to 100\% in our experiments. Simultaneously applied swept microwave (MW) irradiation (with Rabi frequency $\xO_e\lesssim\xo_L$) causes the transfer of polarization to $\Cs$ nuclei in the surrounding lattice. As described in \zfr{theory}A the frequency-swept MW is applied in a sawtooth pattern for $\sim$60s. The DNP occurs \I{independent} of the orientation of the NV center axis in each crystallite, allowing the hyperpolarization of the entire high surface area powder~\cite{Ajoy17}. We evaluate the obtained hyperpolarization by benchmarking the polarization enhancement against the room temperature thermal equilibrium signal at 7T by sample shuttling (see \zfr{theory}A). We note that the sample shuttling time ($\app$648ms) is longer than the electron $T_{1e}$, and to a good approximation therefore the NV electrons are unpolarized during NMR readout. The shuttling time is however small compared to the nuclear $\Cs$ lifetime, $T_{1n}>$120s at $B>$100mT. \zfr{theory}B details the typical dependence on MW sweep rate $\dxo$ and Rabi frequency $\xO_e$, both of which have optimal values set by adiabaticity constraints of the underlying microscopic DNP mechanism. 

To intuitively understand the main features of the DNP mechanism, let us first consider the energy level structure of an NV center coupled to a single $\Cs$ nuclear spin (\zfr{theory}C and \zfr{mechanism}). We work at low fields where $\xo_L<|A|$, and the dominant nuclear quantization axis in the $m_s=\pm 1$ manifolds are set by the hyperfine coupling, referred to as $\xb_{\up,\dw}$ in \zfr{theory}C. The $m_s=0$ state is magnetically silent, and in that manifold the nuclear eigenstates $\xa_{\up,\dw}$ are dominated (for weakly coupled $\Cs$) by the Zeeman field with a second order correction from the hyperfine field, $\wt{\xo}_L \app \xo_L + \fr{\xg_e BA\sin\xt}{\xD-\xg_e B\cos\xt}$, where $\xD$=2.87GHz is the zero-field splitting and  $\xt$ is the angle from the applied field to the N-V axis (\zfr{theory}D). Given our detected $\Cs$ linewidths ($<$1kHz), it is these relatively weakly coupled $\Cs$ nuclei, 300kHz$\:\lesssim|A|\lesssim\:$1MHz that participate most strongly in the hyperpolarization process~\cite{Ajoy17} (see also Supplemental Information~\cite{SOM}). 

Crucially, the large separation between the nuclear eigenstates in the $m_s=\pm$1 manifolds and the low MW powers employed ensure that the swept MWs \I{sequentially} excite a set of transitions that drive the polarization transfer. This manifests as a pair of LZ crossings in the rotating frame. \zfr{mechanism} (reproduced from~\cite{Ajoy17}) shows this for this in the $m_s$=-1 manifold considering positive and negative $A_{zz}$ hyperfine couplings, where crossings occur between the states $\ket{0,\xa_{\up}}\leftrightarrow\ket{-1,\xb_{\up}}$ and $\ket{0,\xa_{\dw}}\leftrightarrow\ket{-1,\xb_{\dw}}$. Under the condition that one is adiabatic with respect to the larger energy gap and positive $A_{zz}$ (see \zfr{mechanism}A), traversal through the level-anticrossings leads to a complete (bifurcated) transfer of population starting from the states $\ket{0,\xa_{\up(\dw)}}$, causing a bias in the system that hypepolarizes the nuclei to the state $\xa_{\up}$. This simple model also captures why experimentally we find that the $\Cs$ DNP sign depends only on the \I{direction} of the microwave sweep, hyperpolarized aligned (anti-aligned) to $B$ under microwave sweeps from low-to-high (high-to-low) frequencies~\cite{Ajoy17}. While the laser is applied simultaneously with the MW sweep, it is of sufficiently low power that the optical repolarization of the NV takes place far away from the LZ events, predominantly during the longer intervals separating  successive sweeps - the laser serving just to reset the NV center to the $m_s=0$ state. The optimal sweep rates (see \zfr{theory}B) are set by adiabaticity constraints that maximize the differential LZ transfer probability between the two pairs of level-anticrossings, and depend both on the Rabi frequency as well as on the hyperfine coupling and orientation (see \cite{SOM}). A more detailed description of the exact energy gaps, and numerical evaluation of the adiabaticity requirements will be presented in a forthcoming publication~\cite{Zangara18}.

While considering the more realistic scenario of multiple $\Cs$ nuclei coupled to the NV center, one obtains a continuum of levels stemming from the hierarchy of the hyperfine interactions~\cite{Dreau14}, the closeby $\Cs$'s dominating the spectral widths. The density of states reflect the underlying hyerfine-broadened electron linewidth (see \zfr{lw}A).  Sweeping over any small spectral window in the broadened line (\zfr{theory}C) still leads to hyperpolarization the sign of which depends on the direction of sweep (see \zfr{results}A).

Even with this brief description, it is already apparent why the hyperpolarization is inefficient with a single sweeper. The electron resonance frequencies $\xD\pm \xg_eB\cos\xt$ are orientation dependent and in a randomly oriented powder the ESR spectrum is broadened to $\mB=2\xg_eB\app$ 1.12GHz at 20mT. This is shown in \zfr{results}A for example, where we indirectly map the NV center ESR spectrum at 27.7mT from the $\Cs$ hyperpolarization enhancement by performing DNP over small (100MHz) windows swept across in frequency space. The obtained spectrum is a convolution of the ESR spectrum with the employed sweep window, the two extremities of the spectrum correspond to the zero degree orientations. The experiment in \zfr{results}A was performed on a collection of $\app$300 diamond microparticles (Element6) of 200$\mu$m size containing a natural abundance (1.1\%) $\Cs$ and $\app$1ppm of NV centers. Since both the $m_s=-1$ and $m_s=+1$ manifolds contain all the NV center electron packets, it is sufficient to just sweep over one of them to obtain the optimal hyperpolarization on the $\Cs$ nuclei. However, the sweep widths required in the $m_s=-1$ manifold, spanning the 0$^{\circ}$ (at frequency $f_0=\xD\mp\xg_e B$) and 90$^{\circ}$  ($f_{90}=  \fr{1}{2}[\xD + \sqrt{\xD^2 + (2\xg_e B)^2}]$) NV center orientations are still rather large (614MHz at 20mT). 
Due to fixed sweep rates $\dxo$ constrained by adiabaticity, the large $\mB$ leads to a long MW sweep time $T=\mB/\dxo\app$16ms$>\trepol$ that far exceeds the repolarization time and bottlenecks the DNP enhancement. Given the laser power employed in our experiments $\app$80mW/mm$^2$, we estimate $\trepol\sim$1ms, on the same order as $T_{1e}$~\cite{Jarmola12}. The exact $\trepol$ is challenging to measure especially on account of optical scattering, total internal reflections, and NV center charge dynamics. 

\begin{figure}[t]
  \centering
  \includegraphics[width=0.5\textwidth]{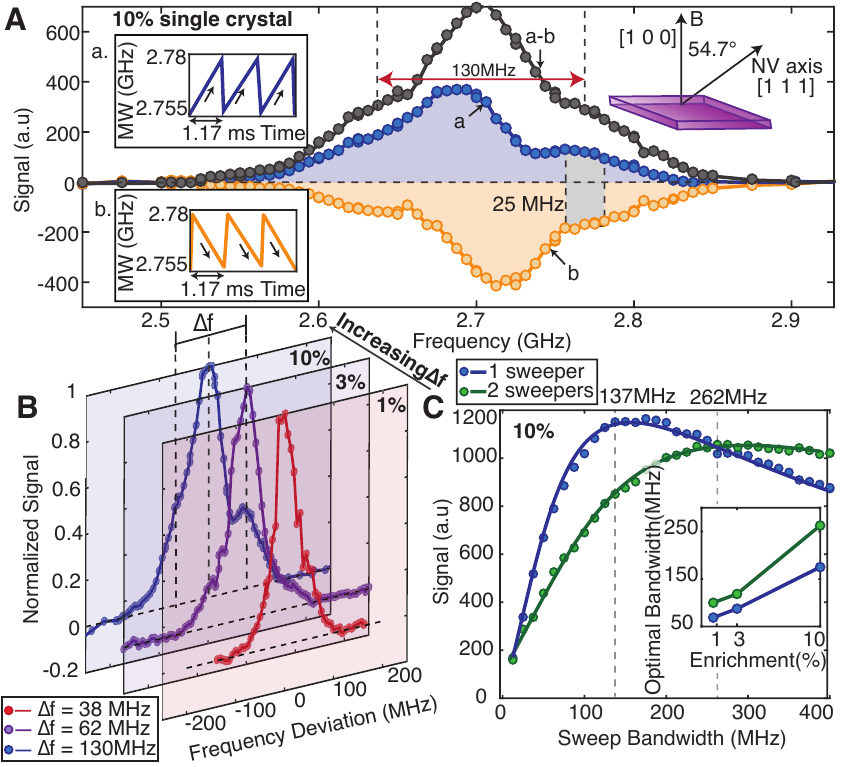}
  \caption{\textbf{ESR linewidth limits to enhancement gain} elucidated by performing $\Cs$ DNP on single crystals with [100] axis placed parallel to the field $B$. All NV axes are then at the magic angle $\xt_M$=54.7$^{\circ}$ to $B$ (inset). (A) \I{ESR lineshape} of a 10\% enriched $\Cs$ crystal mapped indirectly via DNP on a 25MHz window at $B$=10.5mT. Blue (yellow) points show obtained DNP enhancements sweeping MWs from low-to-high (high-to-low) frequency with $\dxo=21$MHz/ms (insets show schematic ramps for an exemplary window). Hyperpolarization sign depends on the direction of MW sweep, and the mirror symmetry of the lineshapes reflect the local density of states (\zfr{theory}C). Black line is the difference signal and faithfully represents the hyperfine broadened NV center ESR spectrum with $\xD f$ dominated by coupling to first shell $\Cs$ nuclei, hyperfine coupled by $\sim$ 130MHz. (B) \I{Effects of $\Cs$ enrichment.} Indirectly mapped ESR spectra of 1\% , 3\% and 10\% $\Cs$ enriched single crystals under DNP with low-to-high frequency sweeps, showing increasing $\xD f$ (legend: FWHM) with enrichment. (C) \I{Cascaded sweeps over a hyperfine broadened line.} For the 10\% $\Cs$ single crystal in (A), we perform DNP with one and two cascaded sweepers over a varying bandwidth centered at 2.688 GHz (peak of (a) in A). Results indicate that frequency comb teeth have to be separated beyond $\xD f$ to provide enhancement gains. \I{Inset:} Optimal sweep bandwidths while employing one and two cascaded sweepers for crystals of different $\Cs$ enrichment. }
\zfl{lw}	
\end{figure}

\T{\I{Swept frequency combs for multiplicative DNP gain:}} --  Frequency combs provide an elegant means to overcome these bottlenecks,
\I{decoupling} the rate at which the NV centers are swept over, and the effective rates at which the LZ anti-crossings are traversed for polarization transfer to the $\Cs$ nuclei. Indeed, a swept microwave frequency comb can maintain the adiabaticity constraints for a single sweep while increasing the cumulative number of sweeps in the total DNP period bounded by nuclear relaxation time $T_{1n}$.  Moreover, in experiments where the NV repolarization rate $\trepol\sim T_{1e}$, the swept frequency comb can ensure that the NV electrons are swept over sufficiently slowly so as to maximize the NV electron polarization at every sweep event. 

Microwave frequency combs can be constructed by semiconductor lasers under negative optoelectronic feedback~\cite{Juan09,Chan07} and nonlinear mixing in tunneling junctions~\cite{Hagmann11}. In this paper we follow a more brute-force approach instead, time-cascading MW sweeps generated by $N$ voltage controlled oscillator (VCO) sources~\cite{SOM}. \zfr{results}B shows the effect of employing a frequency comb for DNP in the NV-$\Cs$ system. The DNP enhancement gains are significant, scaling linearly with $N$, and allowing a multiplicative boost to the DNP enhancement over 7T from 30x to 100x. This constitutes an order of magnitude decrease in averaging time for the same SNR. In the experiments, all the sources sweep the entire bandwidth $\mB$, and the frequency ramps are time-shifted by $\mB/(N\dxo)$ so as to maximize the period between successive sweeps (\zfr{lac}B). While one could consider an alternate scenario where $\mB$ is partitioned into $N$ sub-bands which are swept by individual sources, the current strategy performs better since the electrons at the boundaries between the frequency partitions are also swept across completely, as required for optimal LZ population transfer. Certain implementational aspects deserve brief mention. The individual VCOs have slightly different frequency-voltage characteristics, and to cascade them effectively we \I{match} their exact sweep bandwidths to within $\xD f<1$MHz via a fast-feedback algorithm (see Supplementary Information~\cite{SOM}). The sources are then power combined and amplified, with the amplifier operating well below compression to prevent inter-modulation distortion (IMD) artifacts~\cite{SOM}. 

The cascaded sweeps entail an increase of the total microwave power seen by the sample. For DNP mechanisms (eg. ISE) where the energy gap (see \zfr{lac}) is predominantly determined by the electron Rabi frequency, employing a higher MW power leads to a faster $\dxo$, and the same gains in principle can be achieved by the use of a single sweeper with higher power.
However even in this case, there are several technological advantages of using swept frequency combs for DNP. The costs of MW sources and amplifiers scale rapidly ($\app$ quadratically) with power~\cite{Joye06,Denysenkov08}, but employing a cascade of $N$ low-power amplifiers leads to only a linear cost scaling. Moreover it is easier to directly synthesize slower frequency sweeps~\cite{Guy15}, for instance using inexpensive AWGs and mixers (see Supplemental Information~\cite{SOM}). Our frequency comb method allows one to harness several slow, low-power, sweeps to gain the advantages of more expensive high-power platforms, an advantage especially pertinent at high fields. Moreover, the technique highlights the inherent merits of frequency swept modalities as opposed to field swept ones; while they are equivalent for a single sweep~\cite{Henstra88b}, when cascaded into swept frequency combs the former can provide multifold DNP gains.


\T{\I{Limits of multiplicative DNP gains:}} -- Let us finally evaluate the factors affecting the ultimate limits to the multiplicative enhancement gain. In \zfr{results}C, for a fixed $\dxo$, we vary the sweep bandwidth, equivalent to bringing the frequency comb sweeps closer in frequency and time. We measure the multiplicative gain by normalizing the signal of $N$ sweepers against that from a single one. We observe that when the frequency comb teeth are separated by under $\app$50MHz, there is first a saturation in the DNP boosts and subsequent drop. We ascribe this to the inherent limit set by hyperfine mediated electron broadening $\xD f$ in the powder pattern (see also \zfr{lw}). $\xD f$ here being the \I{width} of each individual NV center electron packet. When two sweeps occur simultaneously on different parts of the ESR line corresponding to a single NV center (\zfr{theory}C), there is interference between them and consequently lower efficiency in the hyperpolarization transfer. While we do expect that the comb teeth separation is ultimately also limited by Rabi frequency, in our experiments $\xO_e\app$430kHz, and it does not play a significant role in setting the limits in \zfr{results}C.

Similar experiments allow us to quantify the optimal spacing between successive comb teeth. In \zfr{phase} we change the \I{phase} $\xph$ between the time-cascaded ramps (e.g. in \zfr{lac}B) that generate the swept frequency combs. More intuitively, this phase directly corresponds to the frequency separation between the successive comb teeth as indicated in the lower panel of \zfr{phase}. Note that the phase of the MWs are arbitrary, here we refer to the phase of the sawtooth patterns generating the frequency sweeps. Changing phase has the effect of varying the frequency comb teeth separation over the fixed sweep bandwidth by $\xph\mB/(2\pi)$ and the time period between successive electron sweeps by $\xph\mB/(2\pi\dxo)$. Intuitively, one would expect that the $\trepol$ and $\xD f$ limits would require that the comb teeth be \I{maximally} separated in both frequency and time, entailing a frequency separation of $\mB/N$, and phase separation $\xph_{\R{opt}}=2\pi/N$. \zfr{phase} confirms this simple picture. Interestingly, it also demonstrates how the enhancement gains arise from the use of multiple sweepers. When all ramps have the same phase, the enhancement from the comb is identical to that employing a single sweeper (dashed line in \zfr{phase}), increasing as the ramps are phase-shifted, with the expected optimal DNP gains at phase separation $\xph_{\R{opt}}$. The plateaus in \zfr{phase} indicate that the enhancement gains are achievable as long as the comb teeth are separated beyond $\xD f$. 

To make this more concrete, in \zfr{lw} we perform DNP on single crystals with different $\Cs$ enrichment. The crystals have $\app$1ppm NV centers, and since we are in the dilute electron limit, the electron packet linewidths are dominated by couplings to the $\Cs$ nuclei. Moreover, the crystals are oriented parallel to the [100] direction such that all N-V axes are equivalent and at the magic angle to the polarizing field $B$ and have the same frequency, hence eliminating inhomogeneous broadening.  This is most evident in \zfr{lw}A that demonstrates the electron spectrum mapped via $\Cs$ DNP, evidenced by the mirror symmetry in the obtained DNP signals with opposite sweep directions.  We note that while we had considered the theory of the hypepolarization mechanism in the context of weakly coupled $\Cs$ nuclei that contribute predominantly to the bulk NMR signal that we measure, \zfr{lw}A also provides direct insight into strongly coupled first shell nuclei. The asymmetry in the obtained ESR spectra directly reports on the polarization of the first shell $\Cs$ spins. Indeed, the difference signal obtained from alternate sweep directions (black line in \zfr{lw}A) shows the characteristic ESR spectrum with satellites from first-shell $\Cs$ nuclei strongly hyperfine coupled by $\sim$130MHz~\cite{Rao16,Parker17}. 
 
 The fact that hyperfine couplings dominate the ESR linewidths (\zfr{theory}C) are most evident in \zfr{lw}B, where we measure $\xD f$ with increasing $\Cs$ enrichment. This is in contrast to \zfr{results}A where the ESR spectrum was inhomogenously broadened due to different orientations of the NV centers in a random powder. In \zfr{lw}C we study the DNP enhancements with one and two cascaded sweepers for varying sweep bandwidths and fixed $\dxo$ over these hyperfine broadened lines, choosing as a representative example the 10\% enriched sample studied in \zfr{lw}A. The sweep bandwidths in these experiments are centered at the peak of the ESR spectrum. Let us first consider the case of a single sweeper (blue line in \zfr{lw}C). The enhancement increases with sweep bandwidth, reaching an optimal value when $\mB\app\xD f$, corresponding to the MWs being applied most efficiently over the electron spectrum. While employing two sweepers on the other hand, there are no DNP enhancement gains when the comb frequencies are closer than $\xD f$. Note that in these experiments we space the comb teeth (optimally) by half the bandwidth. The maximum enhancement occurs when the comb separation is $\xD f$, corresponding to a total sweep bandwidth of $2\xD f$, a strong indication that the two sweepers interfere with each other when simultaneously employed on the hyperfine broadened electron line. Performing similar experiments on the samples with different $\Cs$ enrichment allows us to quantify the sweep bandwidths at which two sweepers perform better than a single one (see Supplemental Information \zfr{powder_bandwidth}). The optimal sweep bandwidths for one and two sweepers are elucidated in the inset of \zfr{lw}C, and they closely match the intrinsic $\xD f$ linewidths in \zfr{lw}B,scaling with $\Cs$ enrichment. Overall \zfr{phase} and \zfr{lw} demonstrate the inherent constraints of the technique, and in the ultimate limit, the frequency combs approach an excitation of all $\xD f$-wide electron packets at once, sweeping them as often as $\trepol$ -- approaching the efficiency of a pulsed DNP experiment over the entire electron bandwidth.

\T{\I{Applications to conventional DNP:}} -- Let us now describe how the current method can be applied in the context of DNP with electron radicals. The most direct applications are for DNP in systems with endogenous radicals (eg. Si, diamond surfaces~\cite{Cassidy13}) or where they can be optically excited~\cite{Capozzi17}, since the linewidths of such radicals are hard to control. 
For radicals with a large \I{g}-anisotropy (e.g. TEMPO, Galvinoxyl) and low-concentrations, the large inhomogeneous broadening leads to a inefficient transfer and lower DNP enhancement due to the (differential) solid effect.  Indeed, experiments are typically performed at higher radical concentrations ($>$20mM~\cite{Siaw14}), where DNP can occur via the cross effect~\cite{Hwang67}, with significantly faster growth times.  However the higher radical concentrations lead to a broadening of the observed NMR lines~\cite{Lange12} and are a challenge for high-resolution spectroscopy applications~\cite{Saliba17}. The use of swept frequency combs can dramatically improve DNP enhancement at low radical concentrations by enacting a transition to the integrated solid effect~\cite{Dirksen89}. Since ISE is bottlenecked by similar factors, the demonstrated gains in \zfr{results} should be directly transferable to ISE. The use of frequency modulation to implement ISE was recently demonstrated at X-band~\cite{Can17}, and we anticipate large gains with frequency combs. While the increased reliance on spin diffusion will increase the growth time of the DNP signal, frequency combs offer the possibility of obtaining high DNP enhancements without concomitant NMR line broadening. 

\T{\I{Conclusions and outlook:}} -- We have proposed and experimentally demonstrated a simple and scalable technique to obtain multiplicative enhancement gains in dynamic nuclear polarization. The method entails a swept frequency comb to excite the entire inhomogenously broadened electron bandwidth for polarization transfer. It can be implemented by cascading $N$ sweeps from individual low-power sources/amplifiers to obtain a DNP enhancement boost $\propto N$, with ultimate limits set by the hyperfine-mediated electron linewidth and lifetime $T_{1e}$. As such the technique affirms the notion the electron spin control can significantly enhance DNP by harnessing the full power of the electron spectrum.  We demonstrated its utility for the hyperpolarization of $\Cs$ nuclei in diamond microparticles via optically pumped NV centers at room temperature, obtaining a 300\% boost in DNP efficiency. When employed for conventional polarizing radicals at high fields, the technique promises to yield DNP enhancement boosts in excess of one order of magnitude, with a relatively simple implementation employing existing technology and only modest cost overheads.

\T{\I{Acknowledgments:}} -- We gratefully acknowledge discussions with J.P. King, P. R. Zangara and S. Dhomkar. CAM acknowledges support from the National Science Foundation through grants NSF-1401632 and NSF-1547830 and from Research Corporation for Science Advancement through a FRED Award. CR acknowledges support from the National Science Foundation through grant CHE-1410504.

\bibliography{C:/paper-drafts/Biblio}
\bibliographystyle{apsrev4-1}

\pagebreak
\clearpage
\onecolumngrid
\begin{center}
\textbf{\large{\textit{Supplementary Information:} \\
\bluetitle{Enhanced dynamic nuclear polarization via swept microwave frequency combs}}}\\
\hfill \break
\smallskip
\begin{minipage}[t]{0.625\textwidth}
\begin{center}
\end{center}
\end{minipage}
\end{center}

\twocolumngrid

\beginsupplement
{ \hypersetup{linkcolor=darkred}
\tableofcontents
}

\begin{figure}[t]
  \centering
  {\includegraphics[width=0.45\textwidth]{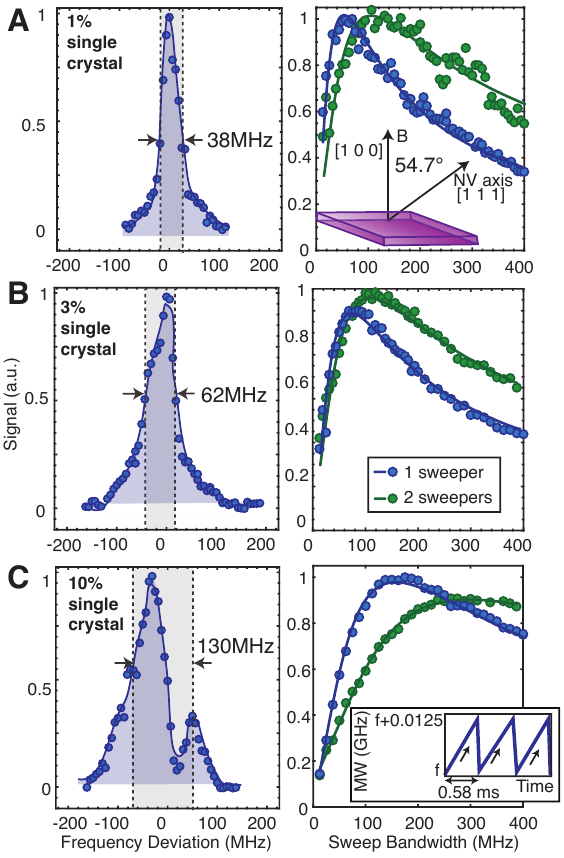}}
  \caption{
    \textbf{Linewidth limits of multiplicative DNP gains.} \I{Left panels:} NV center electronic spectrum mapped indirectly via $\Cs$ DNP in a 12.5MHz window for single crystals of (A) 1\% (B) 3\% and (C) 10\% $\Cs$ enrichment. Crystals are placed parallel to the [100] axis (inset), so that all families of NV centers have overlapping spectra. Blue data indicate the obtained $\Cs$ enhancements sweeping microwaves from low-to-high frequency (\I{inset}). Solid lines are guides to the eye. Electron linewidth $\xD f$ (FWHMs are marked) increases with enrichment and the sign of $\Cs$ hyperpolarization only depends on the direction of MW sweep. \I{Right panels:} Obtained DNP enhancements employing one (blue) and two (green) sweepers for varying bandwidths over the electron spectrum. Sweep bandwidths are centered on the spectra in the left panels. Multiple sweeper combs provide an advantage over a single sweeper when the comb teeth are separated by more than the electron linewidth.
}
\zfl{single-crystals}
\end{figure}

\begin{figure}[t]
  \centering
  {\includegraphics[width=0.45\textwidth]{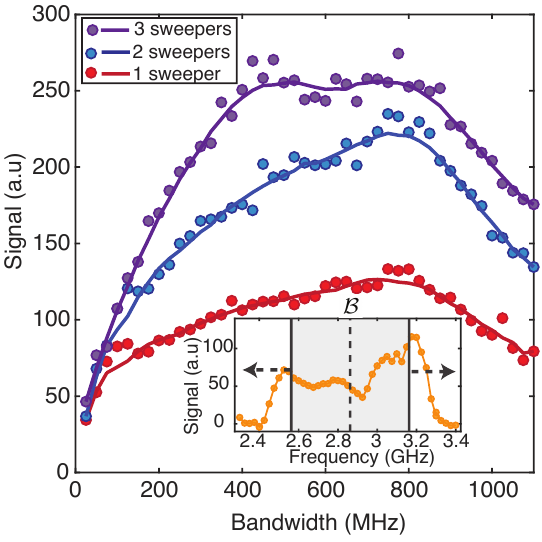}}
  \caption{
    \textbf{Bandwidth dependence of multiplicative DNP gains} in microdiamond powder. For varying sweep bandwidths centered on the NV center powder pattern, we obtain the $\Cs$ DNP signal employing a swept frequency comb of upto three cascaded sweepers at $\app$20mT. Solid lines are guides to the eye.  \I{Inset:} Experimentally determined powder pattern. Shaded region denotes an example 600MHz sweep bandwidth. Panel demonstrates electron linewidth limits on the obtained multiplicative DNP gain (see \zfr{single-crystals}).
}
\zfl{powder_bandwidth}
\end{figure}

\section{Linewidth limits of frequency comb DNP}
In order to highlight the  linewidth limitations when employing multiple cascaded sweepers, we performed detailed experiments mapping the NV center ESR spectrum via $\Cs$ DNP in a narrow 12.5MHz sweep window for samples of different $\Cs$ enrichments (see \zfr{single-crystals}). Similar to \zfr{lw} in the main paper, the crystals are placed with the NV axes at the magic angle to the polarizing field, allowing the obtained spectra to be broadened predominantly by hyperfine couplings to $\Cs$ nuclei. We ascribe the slight asymmetry in the lineshape of  \zfr{single-crystals}B to crystal misalignment from the magic angle. The linewidth of the highly enriched samples are dominated by the hyperfine coupling to closeby $\Cs$. A systematic discussion of the changes in lineshape is beyond the scope of this work and will be addressed in a forthcoming publication.

It is evident that the two cascaded sweepers provide a gain over a single one (right panels in \zfr{single-crystals}) only when the comb teeth are separated by approximately the hyperfine mediated electron linewidth, which increases with $\Cs$ enrichment. This data was also used to report the exact optimal sweep bandwidths for one and two sweepers, plotted in the inset of \zfr{lw}C of the main paper.

\zfr{powder_bandwidth} studies a similar bandwidth dependence of DNP gains with multiple cascaded sweepers for a 1\% (natural abundance) microdiamond sample. The sweep bandwidth is centered at the center of the powder pattern (inset of \zfr{powder_bandwidth}). Once again, the enhancement gains are a strong function of the sweep bandwidth, decreasing sharply when the comb teeth spacing approaches the hyperfine mediated electron linewidth.

\begin{figure*}[t]
  \centering
  {\includegraphics[width=0.9\textwidth]{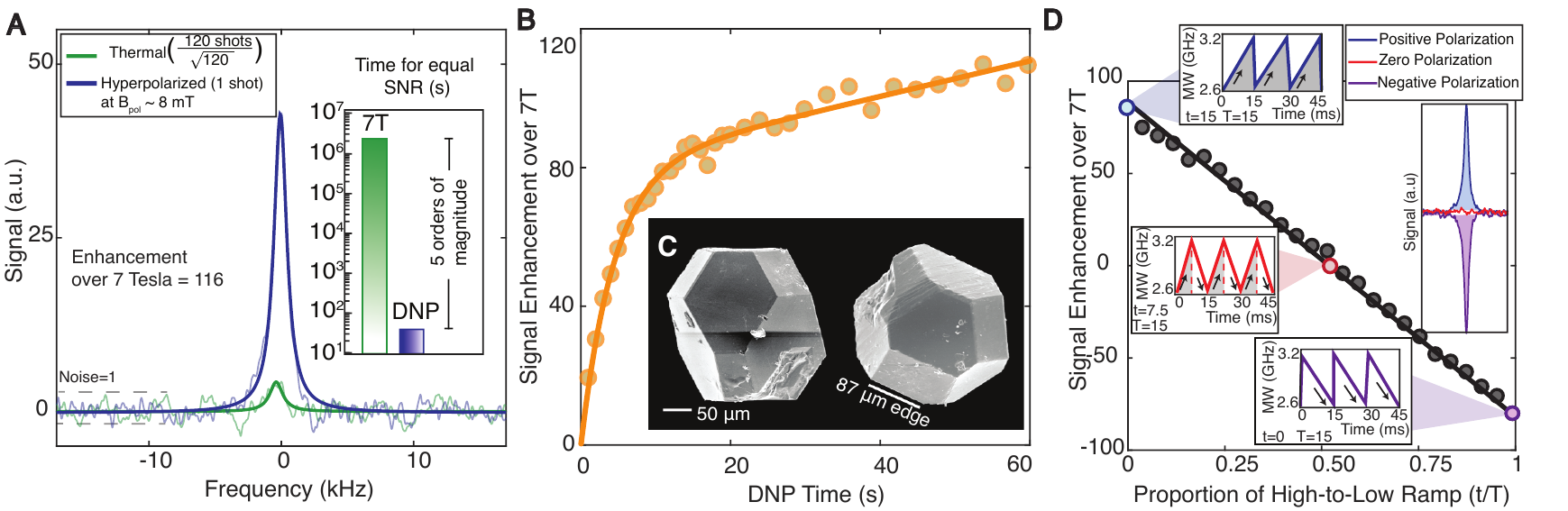}}
  \caption{\T{Optical hyperpolarization in diamond microparticles.} Hyperpolarization experiments were performed on 200$\mu$m HPHT particles~\cite{Ajoy17}. Solid fit lines are depicted over experimental data points. (A) \I{Signal gain by DNP} under optimized conditions. Green line shows the $\Cs$ NMR signal due to Boltzmann polarization at 7T, averaged 120 times over 7 hours. Blue line is a single shot DNP signal obtained with 40s of optical pumping, enhanced by 116 over the 7T thermal signal (enhanced 101500 times at polarizing field $B$=8mT). The signals have their noise unit-normalized for clarity. Hyperpolarization thus leads to over 5 orders of magnitude gains in averaging time (inset). (B) \I{Buildup curve} showing rapid growth of bulk $\Cs$ polarization. Slow rise at longer times is reflective of $\Cs$ spin diffusion. (C)\I{SEM micrographs} (Hitachi S5000) of two individual e6 HPHT diamond particles. The particles have a uniform size distribution (edge length $87\pm3.9\ \mu$m), and a truncated octahedral shape set by particle growth conditions.  (D) \I{Hyperpolarization sign} is controlled by MW sweep direction across the NV center powder pattern. Continuous family of sawtooth-sweeps demonstrating the concept, varying the duty cycle of upward ramps. Extremal points represent low-to-high frequency MW sweeps and vice-versa. \I{Inset:} $\Cs$ signal undergoes near-perfect sign inversion upon reversal of the sweep direction. Sweeping in a symmetric fashion leads to net cancellation, and no buildup of hyperpolarization.}
 \zfl{s1}	
\end{figure*}

\begin{figure}[t]
  \centering
  {\includegraphics[width=0.49\textwidth]{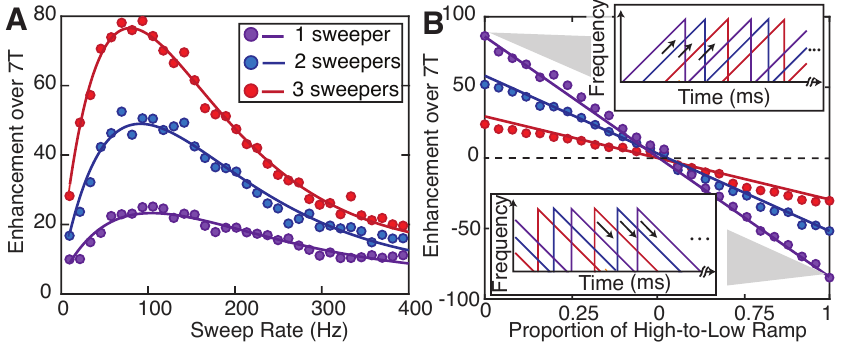}}
  \caption{
     \textbf{Hyperpolarization gains with cascaded sweeps} occurs while maintaining the same sweep rate and polarization direction characteristics of the DNP mechanism. Figure illustrates the (A) sweep rate dependence and (B) sweep direction dependence on the obtained $\Cs$ enhancement while employing upto three cascaded sweepers. The panels indicate that (A) the swept frequency comb provides multiplicative DNP gains while maintaining approximately the same optimal sweep rate, and (B) the hyperpolarization sign that only depends on the direction of the overall microwave sweeps (insets and \zfr{s1}).
}
\zfl{sweep_symm}
\end{figure}

\section{Hyperpolarized signal and buildup}
\zfr{s1} exhibits essential results from Ref. \cite{Ajoy17}, demonstrated for a typical example of 200$\mu$m microparticles with natural abundance $\Cs$ containing about 1ppm of NV centers (\zfr{s1}C). We obtain $\Cs$ hyperpolarization over 116 times that of the 7T Boltzmann level (\zfr{s1}A) with a completely randomly oriented powder. For clarity the signals in \zfr{s1}A have their noise unit-normalized, and a single shot DNP signal has about 10 times the signal-to-noise (SNR) of the 7T thermal signal obtained after $\app$7hr of averaging -- a time gain of over 5 orders of magnitude to get the identical SNR at 7T with thermal signal.

The polarization within the diamond particles builds up to $\app$0.1\% polarization level in about 40s of optical pumping (\zfr{s1}B). This is reflective of the efficiency of the low field mechanism, being able to hyperpolarize a large number of $\Cs$ nuclei (not just in the first shell). The pumping time is limited by the nuclear lifetime $T_{1n}\app 20$s at low field ($\app$ 20mT). The polarization buildup curve exhibits a linear ramp at long times due to slow spin diffusion away from the directly polarized $\Cs$ spins. Increasing $\Cs$ enrichment leads to more rapid spin diffusion and consequently faster DNP buildup.

In our experiments, the sign of the polarization only depends on the direction of the microwave sweeps (\zfr{s1}C). Sweeping the microwaves in a ramp fashion from low-to-high frequency leads to nuclear polarization aligned to the polarizing $B$ which we term positive polarization. Anti-alignment can be achieved accordingly by sweeping from high-to-low-frequency. This allows on-demand control of the sign of polarization. As expected, a triangular sweep pattern with equal amounts of high-to-low and low-to-high frequency sweeps leads to destructive interference in alternate periods, and no net polarization buildup. Indeed, increasing the number of cascaded sweepers $N$ maintains the same optimal sweep rate set by adiabaticity constraints, and hyperpolarization sign dependence on the direction of the sweep (see \zfr{sweep_symm}).

\section{Hyperpolarization Mechanism}
We now briefly describe the low field DNP mechanism that governs the polarization transfer in our experiments. For more details, and experimental characterization of the mechanism, we point the reader to Ref. \cite{Ajoy17}. Consider for simplicity a NV center coupled to a single $\Cs$ nuclear spin. The Hamiltonian of the system is,
\beq\label{eq:1}
\begin{aligned}
	\mH = {} & \xD S_{z}^2 - \xg_e \vec{B} \cdot \vec{S} - \xg_n \vec{B} \cdot \vec{I} + A_{zz}S_{z}I_{z}\\
	& + A_{yy}S_{y}I_{y} + A_{xx}S_{x}I_{x} + A_{xz}S_{x}I_{z} + A_{zx}S_{z}I_{x}
\end{aligned}
\eeq
where $\vec{S}$ and $\vec{I}$ respectively denote the NV and $\Cs$ vector spin operators, and $\vec{B}$ is the magnetic field (10-30 mT) at angle $\theta$ ($\phi$) to the NV axis. Within the $m_s=\pm1$ states, the hyperfine coupling produces a $\Cs$ splitting,
\beq\label{eq:2}
	\xo_{C}^{(\pm1)} = \sqrt{(A_{zz}\mp\xg_{n}B\cos\xt)^2 + A_{zx}^2}
\eeq
For the $m_s=0$ manifold, second-order perturbation theory leads to the approximate formula~\cite{Alvarez15},
\beq\label{eq:3}
\begin{aligned}
	\wt{\xo}_L \approx {} & \xg_{n}B \\
	& + 2\Big(\frac{\xg_{e}B}{\xD}\Big) \sin\xt \big(\sqrt{A_{xx}^2 + A_{zx}^2} \cos^2\phi  + A_{yy}\sin^2\phi \big)
\end{aligned}
\eeq 
From Eqs. \ref{eq:2} and \ref{eq:3} we conclude that each manifold (including the $m_s=0$ manifold) has its own, distinct quantization axis which might be different from the direction of the applied magnetic field. In particular, the second term in Eq. \ref{eq:3} can be dominant for hyperfine couplings as low as 1 MHz (corresponding to nuclei beyond the first two shells around the NV) if $\theta$ is sufficiently large, implying that, in general, $\Cs$ spins coupled to NVs misaligned with the external magnetic field experience a large frequency mismatch with bulk carbons, even if optical excitation makes $m_s=0$ the preferred NV spin state.

Assuming fields in the range 10-30 mT, it follows that $\Cs$ spins moderately coupled to the NV (300 kHz $\lesssim |A_{zz}|\lesssim$ 1 MHz) are dominant in the hyperpolarization process because they more easily spin diffuse into the bulk and contribute most strongly to the observed NMR signal at 7T. For sweep rates near the optimum ($\sim$ 40 MHz/ms), the time necessary to traverse the set of transitions connecting $m_s=0$ with either the $m_s=-1$ or $m_s=+1$ manifolds is relatively short \big($\lesssim$ 30 $\mu$s for weakly coupled carbons\big) meaning that optical repolarization of the NV preferentially takes place during the longer intervals separating two consecutive sweeps, as modeled in \zfr{mechanism}. 

Nuclear spin polarization can be understood as arising from the Landau-Zener crossings in \zfr{mechanism}. Efficient polarization transfer takes place when the narrower LZ crossings connect branches with different electron and nuclear spin quantum numbers, precisely the case in the $m_s=0\leftrightarrow m_s=-1$ ($m_s=0\leftrightarrow m_s=+1$) subset of transitions when the hyperfine coupling is positive (negative). When probing ensembles, both sets of transitions behave in the same way, i.e., $\Cs$ spins polarize positive in one direction, negative in the other. A more detailed exposition of the hyperpolarization mechanism and simulations are presented in Ref. \cite{Ajoy17}.

\begin{figure}[t]
  \centering
  	\includegraphics[width=0.5\textwidth]{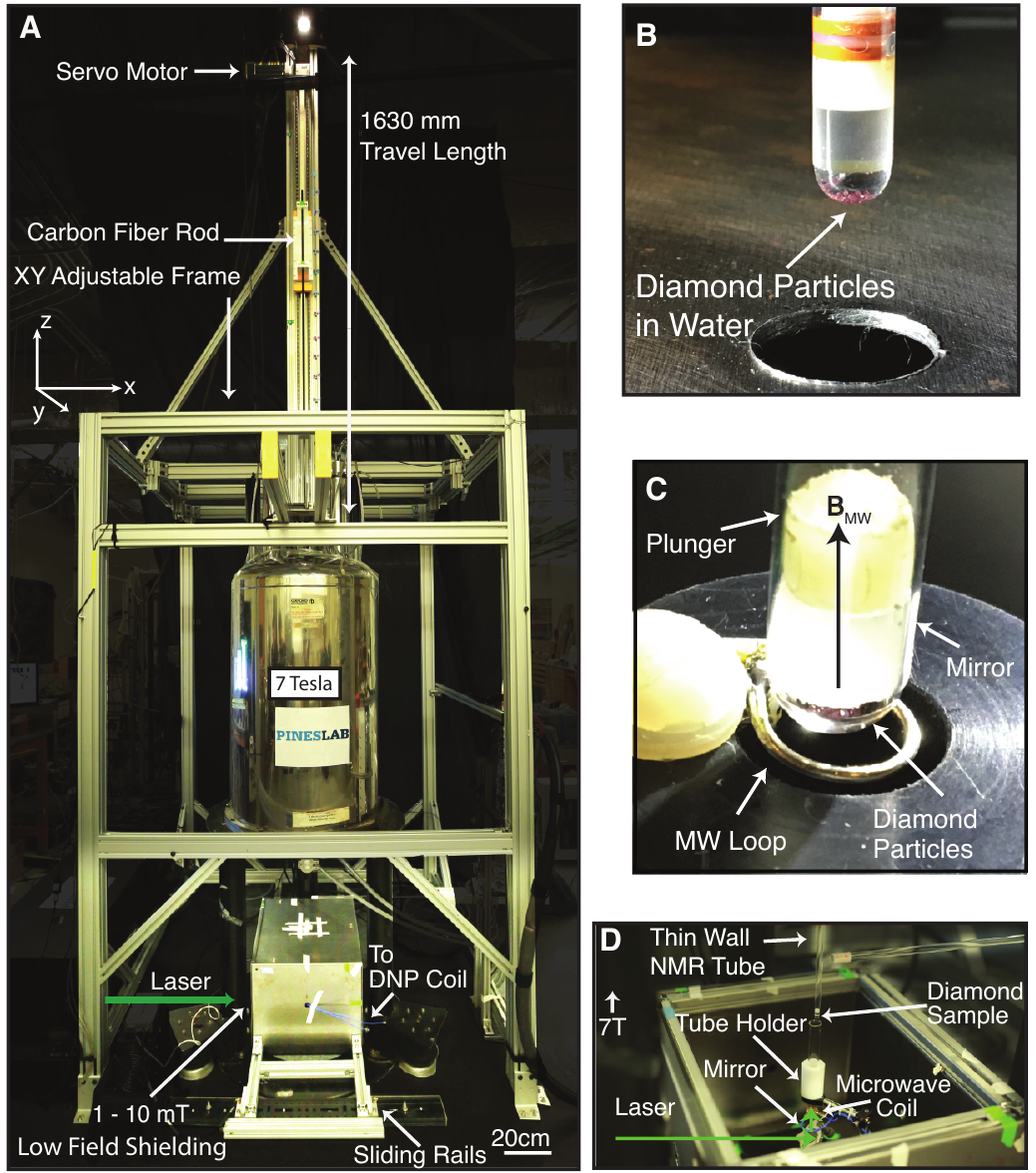}
\caption{\textbf{Hyperpolarization setup.} Hyperpolarization is carried out at low field (1-30mT) and the bulk $\Cs$ polarization measured at 7T by rapid sample shuttling. (A) A mechanical shuttler is constructed at the top of a superconducting magnet (7T) on X and Y adjustable rails for alignment. Hyperpolarization is performed in a low field shield, which is secured on sliding rails. An actuator shuttles a carbon fiber rod with sample tube attached to the end along a conveyer belt for 1630mm travel distance. (B) Diamond particles are distributed in water, and carried in a glass tube. A plunger holds the contained volume firmly to ensure samples stay in position during shuttling. A mirror was employed to concentrate light that is applied from below. (C) Illustration of the MW coil used for exciting the swept frequency combs, consisting of a MW stub antenna.  (D) DNP setup inside the low field shield. Laser irradiation is applied from the bottom of the NMR tube and the tube is positioned above the microwave coil.}
\zfl{shuttler} 
\end{figure}

\begin{figure*}[t]
  \centering
  \includegraphics[width=0.98\textwidth]{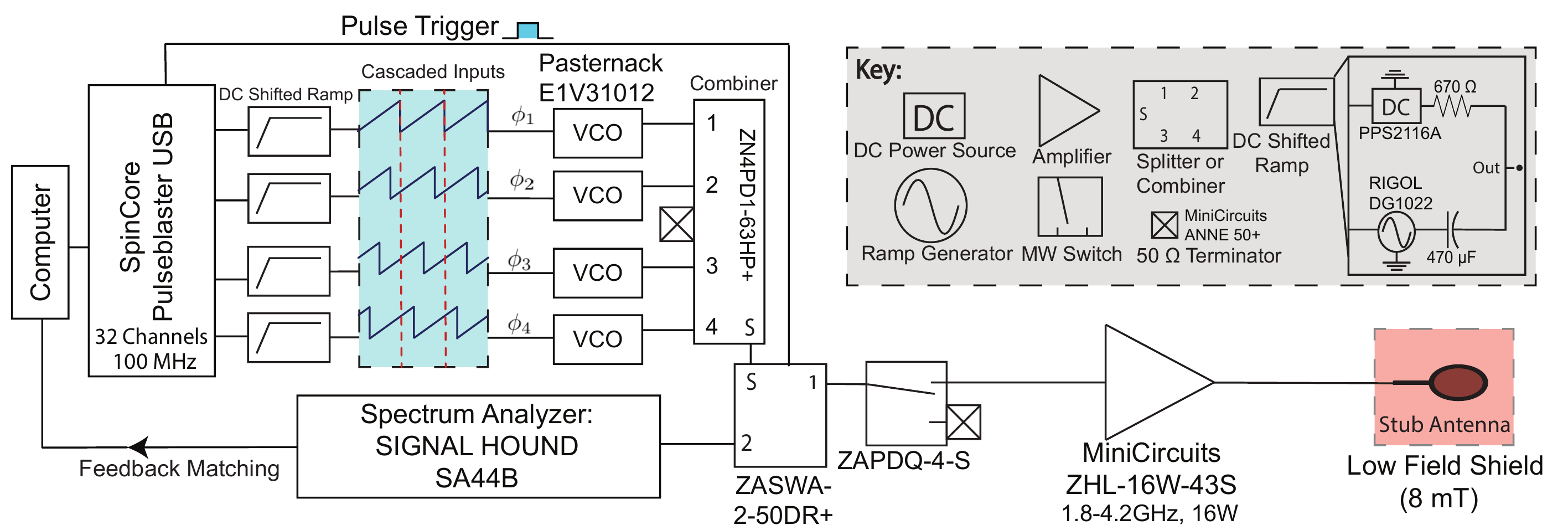}
  \caption{\textbf{Schematic circuit for DNP excitation.} Low field DNP from NV centers to $\Cs$ is excited by microwave sweeps produced by employing voltage controlled oscillators (VCOs) with ramp generator inputs (see \zfr{VCO-matching}). Multiple VCOs are employed in a cascade to increase polarization transfer efficiency. A spectrum analyzer is used to implement a feedback algorithm that exactly matches the VCO bandwidths to $<$2 MHz (see \zfr{VCO-matching2}).  The microwaves are finally amplified by a 16W amplifier into a stub antenna that produces either longitudinal or transverse fields.}
\zfl{DNP-circuit}
\end{figure*}

\begin{figure}[t]
 \centering
  \includegraphics[width=0.44\textwidth]{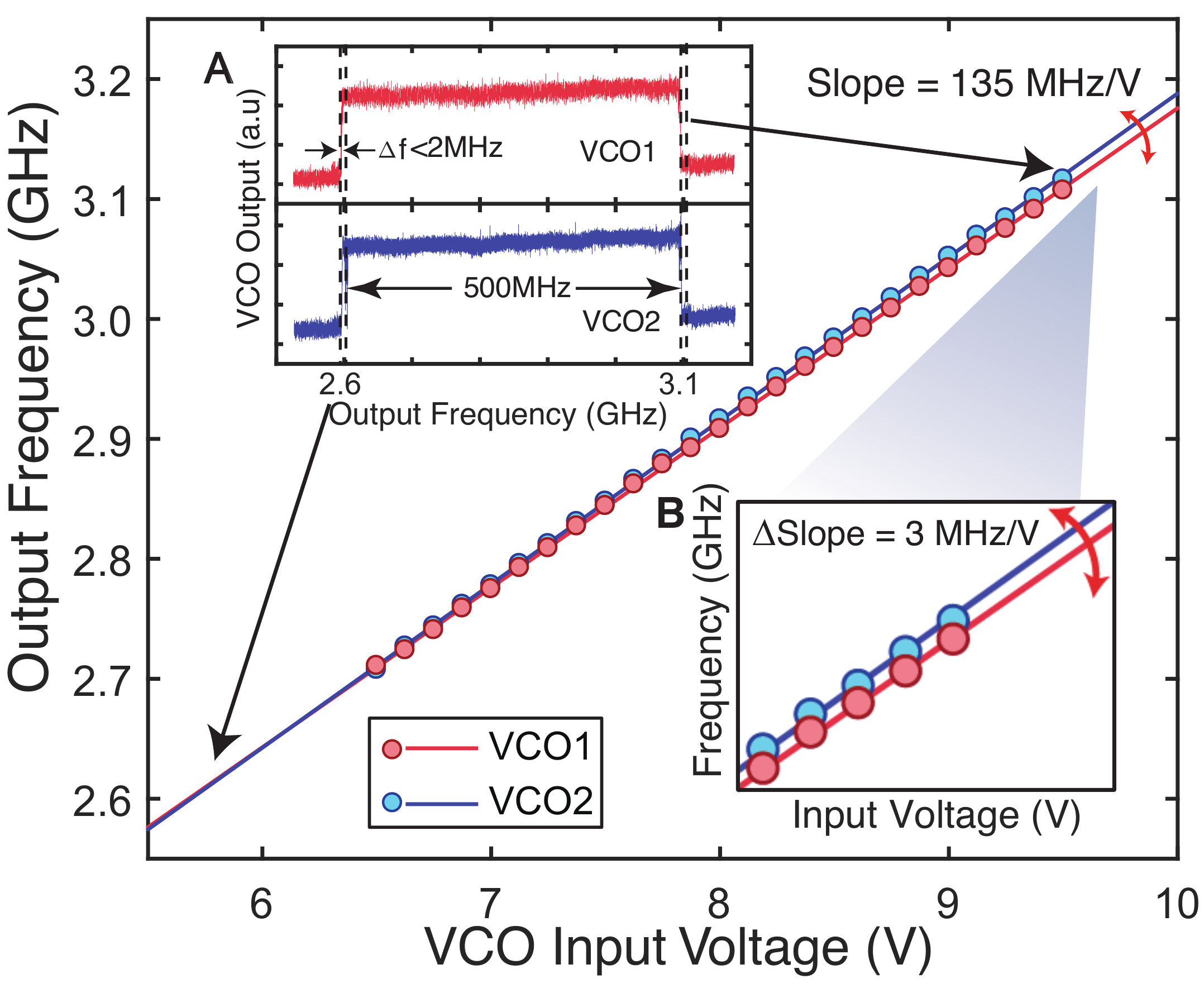}
  \caption{\textbf{Feedback matching of VCO sweep bandwidths.} Frequency-voltage characteristics of two Minicircuits ZX95-3800A+ VCO frequency sources measured with an in-situ spectrum analyzer (\zfr{DNP-circuit}), showing dissimilarity in frequency output with tuning voltage by $\app 3$MHz/V (inset A).  (B) Feedback is now implemented to match their swept bandwidths to 500$\pm$ 1 MHz, shown in the insets for 2kHz sweep frequency (see \zfr{VCO-matching2}).  This allows the VCOs to be cascaded to simultaneously sweep over the NV center powder pattern to enhance DNP efficiency.}
\zfl{VCO-matching}
\end{figure}

\section{Integrated Solid Effect}

For completeness, here we review the principles of Integrated Solid Effect (ISE) (see ~\cite{Henstra88b,Henstra14}) and the adiabaticity conditions for optimal polarization transfer (see Fig. 1 of main paper). The DNP mechanism we employ for the NV-$\Cs$ system shares several implementational similarities with ISE. Under swept MW irradation with frequency $\xo$, the Hamiltonian of a coupled electron nuclear spin system is, 
\beq
\mH=\omega_{e}S_z - \omega_{L}I_z + 2\xO_e S_x\cos(\omega t) + \bm{S}\cdot \bm{A} \cdot \bm{I}.
\eeq 
The first two terms are electron and nuclear Zeeman term respectively, and the the last term is hyperfine coupling between the electron and nuclei. In the rotating frame,
\beq
H=\xD\xo S_z + \xO_e S_{x}-\omega I_z+ S_z A_{zz} I_z +\frac{1}{2} S_z A_{z+} I_{-}+\frac{1}{2} S_z A_{z-} I_{+}
\eeq
where $\xD\xo=\omega_{e}-\omega$, $I_{\pm}=I_x\pm I_y$, $A_{z\pm}=A_{zx}\pm A_{zy}$. Moving to a tilted reference frame via a rotation about $y$-axis, $\exp(i\theta S_{y})$ so that the effective frequency vector ($\xO_e$, 0 , $\xD\xo$) is along the new $z^t$-axis gives,
\beq
\mathcal { { H } }^ { t } = Z ^ { t } + \mathcal { { V } ^ { \dagger }} \sin \theta 
\eeq
where $Z ^ { t } = \omega _ { \text{eff} } S _ { z ^ { t } } - \omega _ {L} I _ { z } ^ { i }$ is the Zeeman part and  ${ V ^ { \dagger } } = - \frac { 1} { 4} ( A _ { z + } ^ { i } S _ { - } I _ { - } ^ { i } + A _ { z - } ^ { i } S _ { + } I _ { + } ^ { i } ) $, assuming that the components of the hyperfine tensors are small. In this tilted rotating frame set by basis $\ket{m_s^t}\ket{m_I}$, $ \mathcal { { V } ^ { \dagger }} \sin \theta$ contributes to non-diagonal term and induces states transitions. There are then effectively two level anti-crossings (LACs) at which polarization polarization transfer from the electron to nuclei can be affected, given by $\sqrt { (\xD\xo) ^ { 2} + \xO^ { 2}} - \omega _ {L} \approx 0$. Assuming the rate for frequency sweep is $\dot { \omega }$, the probability of electron spin staying at the same state equals is $\exp ( - \frac { \pi \xO_e^ { 2} } { 2| \dot { \omega } | } )$. Thus this results in the adiabatic condition for the frequency sweep $\frac { \pi \xO_e ^ { 2} } { 2| \dot { \omega } | } \gg 1$.

\section{Experimental design}
\zsl{shuttler_low_field}

During our DNP process, 1W laser (520nm LasterTack) light is applied continuously for a fixed time ($\sim $60s) to polarize the NV centers. Simultaneously applied swept microwave (MW) irradiation (MiniCircuits ZHL16W-43S+ 16W) across the NV center spectrum at 1-30mT transfers the polarization to $\Cs$ nuclei (see \zfr{theory}A in the main paper). A mechanical field cycler~\cite{Ajoy17I} then rapidly carries the sample from the low field magnetic shield (NETIC S3-6 alloy 0.062'' thick, Magnetic Shield Corp) where hyperpolarization is excited to a 7T superconducting magnet (see \zfr{shuttler}A) in a total travel time of 648$\pm$2.6ms. Inductive detection of the $\Cs$ NMR signal starts immediately after the sample is in position at high field. The entire hyperpolarization procedure is relatively easy to conduct on account of the low laser and MW powers employed, as well as absolutely no requirement for alignment of diamond samples to the magnetic field. The field cycler consists of a high-precision actuator (Parker HMRB08) with a twin carriage mount carrying a carbon fiber rod (8mm, Rockwest composites) into which the NMR tube (8 mm, Wilmad) containing the sample is pressure fit. Single crystal or powder samples are immersed in water (\zfr{shuttler}B), and a plunger firmly holds the sample solution to prevent changes in sample orientation and position during shuttling (\zfr{shuttler}C). Single crystals have the NV axes oriented at magic angle (see \zfr{lw}) to the polarizing field. \zfr{shuttler}D details the hyperpolarization setup in the low field shield. The laser beam is collimated to a $\app$4mm diameter and irradiated at the bottom of the NMR tube carrying the sample. The microwaves are delivered by means of a stub antenna (loop) employed below the tube.  The motion and subsequent detection is controlled and sequenced with pulse generator (SpinCore PulseBlaster USB 100 MHz) using a high voltage MOSFET switch (Williamette MHVSW-001V-036V). For more details on the device construction and performance we point the reader to Ref.~\cite{Ajoy17I}.

\section{Electronics for swept MW frequency combs}
\subsection{Setup}
Microwave (MW) sweeps are applied across the NV center powder pattern to drive the DNP process (see \zfr{theory}A in the main paper). For the experiments in this work, we employed voltage controlled oscillator (VCO) (Minicircuits ZX95-3800A+ (1.9-3.7 GHz)) sources to generate the frequency sweeps, by employing DC shifted ramp input voltages to produce the sweeps (see \zfr{DNP-circuit}). The ramp is produced using a programmable power supply (Circuit Specialists) that generates a DC voltage V$_{\textrm{dc}}$ that is combined with an AC sawtooth ramp with peak to peak value V$_{\textrm{pp}}$ from an arbitrary waveform generator (Rigol 1022A), in a high-pass configuration with a $\sim$ 1Hz cutoff frequency. The input ramps are programmed to carefully tune the VCO outputs to the target sweep bandwidths corresponding to the NV center powder pattern~\cite{Ajoy17}. Lastly, after being power combiened (MiniCircuits ZN4PD1-63HP+), a 16W amplifier (MiniCircuits ZHL16W-43S+) transmits the microwaves to a stub antenna matched to the diameter of the tube containing the sample~\cite{Ajoy17} to excite the $\Cs$ hyperpolarization.

\subsection{Generating swept frequency combs}
Cascaded sweeps utilizing multiple VCOs ($N_{\R{VCO}}$) are generated by using input voltage ramps that are phase shifted by $2\pi/N_{\R{VCO}}$. The VCO output frequency $f(V)$ and input voltage $V$ has a approximate linear relationship, $f(V)=b\cdot V+F$, where $b$ is a constant coefficient and $F\approx$1.9GHz is the frequency when $V=0$. However, the VCOs have slightly differing f-V characteristics, even when of the same family, due to inter-device variation and temperature dependence (see \zfr{VCO-matching}). In order to \I{match} all the VCOs to sweep the target band, and to generate the equally spaced frequency comb, we implemented a gradient descent feedback algorithm employing a fast spectrum analyzer (SignalHound USB-SA44B).  

Let us define the DC and AC voltage inputs to the VCOs for the $i$th iteration ($i$=1, 2, 3...) to be V$_{\textrm{pp}}^i$ and V$_{\textrm{dc}}^i$.  We define center of the spectrum f$^i$ for the $i$th iteration, and bandwidth of the spectrum $\triangle$f$^i$, while the target spectrum center and width are f$_0$, $\triangle$f$_0$ respectively. Given the linearity of the VCO response, V$_{\textrm{dc}}$ and V$_{\textrm{pp}}$ are predominantly related to f and $\triangle$f respectively. The following equations are applied to update V$_{\textrm{pp}}$ and V$_{\textrm{dc}}$ level for each iteration:
\begin{equation}
\left\{
\begin{aligned} 
V_{\textrm{pp}}^{i+1}&=\frac{\triangle f_0}{\triangle f^i}\cdot V_{\textrm{pp}}^i \\
b^i&=\frac{f^i-F}{V_{dc}^i} \\
V_{\textrm{dc}}^{i+1}&=V_{i}+\frac{f_0-f^i}{b^i}
\end{aligned}
\right.
\end{equation}

During each the feedback loop, V$_{\textrm{pp}}$ is adjusted based on the assumption that bandwidth $\triangle$f is approximately proportional to V$_{\textrm{dc}}$, and V$_{\textrm{dc}}$ shifted to approach the target band center. To ensure VCO receiving reasonable input, initial values are set to be V$_{\textrm{pp}}$=2V and V$_{\textrm{dc}}$=6V. 
based on empirical $b$ and $F$. The pre-set deviation we typically use is 2MHz, which is approximately the VCO output linewidth when input is a single constant voltage. The efficiency of the algorithm is highlighted in \zfr{VCO-matching2}. 

\begin{figure}
  \centering
  {\includegraphics[width=0.49\textwidth]{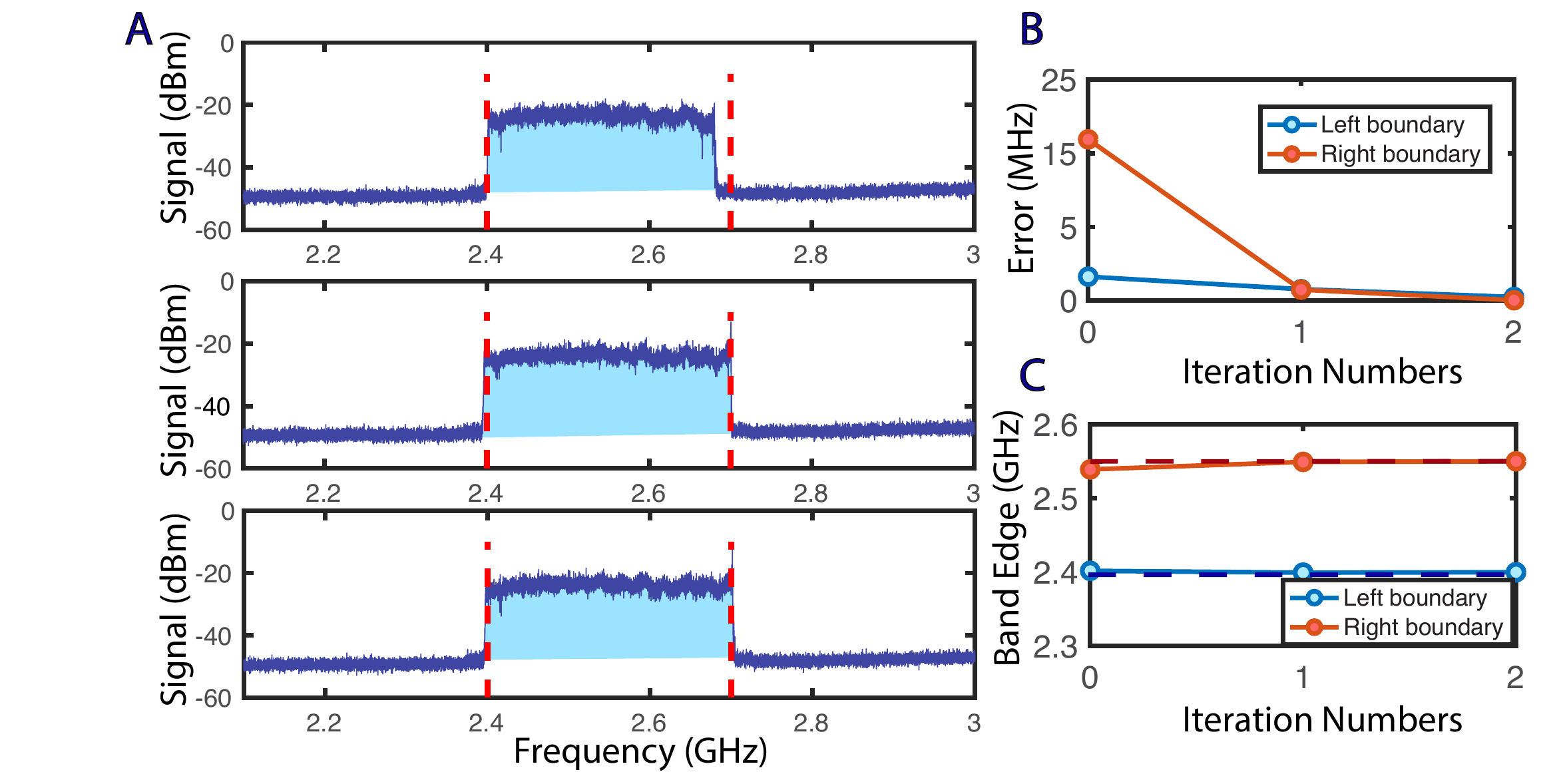}}
  \caption{
   \textbf{Demonstration of VCO matching process}, to a sweep bandwidth of 2.4-2.7GHz, typical for single crystal DNP experiments  in a ~12mT polarizing field field. Only two iterations were taken to match the VCO output to targeted band within deviation of 1MHz. (A) Spectrum of VCO output (2kHz sweep rate) for the individual iterations. (B) Errors to the left and right boundaries of the target sweep band with matching iterations.  (C) Band edges (left and right) converge to the target band in 2 iterations.
}
\zfl{VCO-matching2}
\end{figure}

\begin{figure}
  \centering
  {\includegraphics[width=0.49\textwidth]{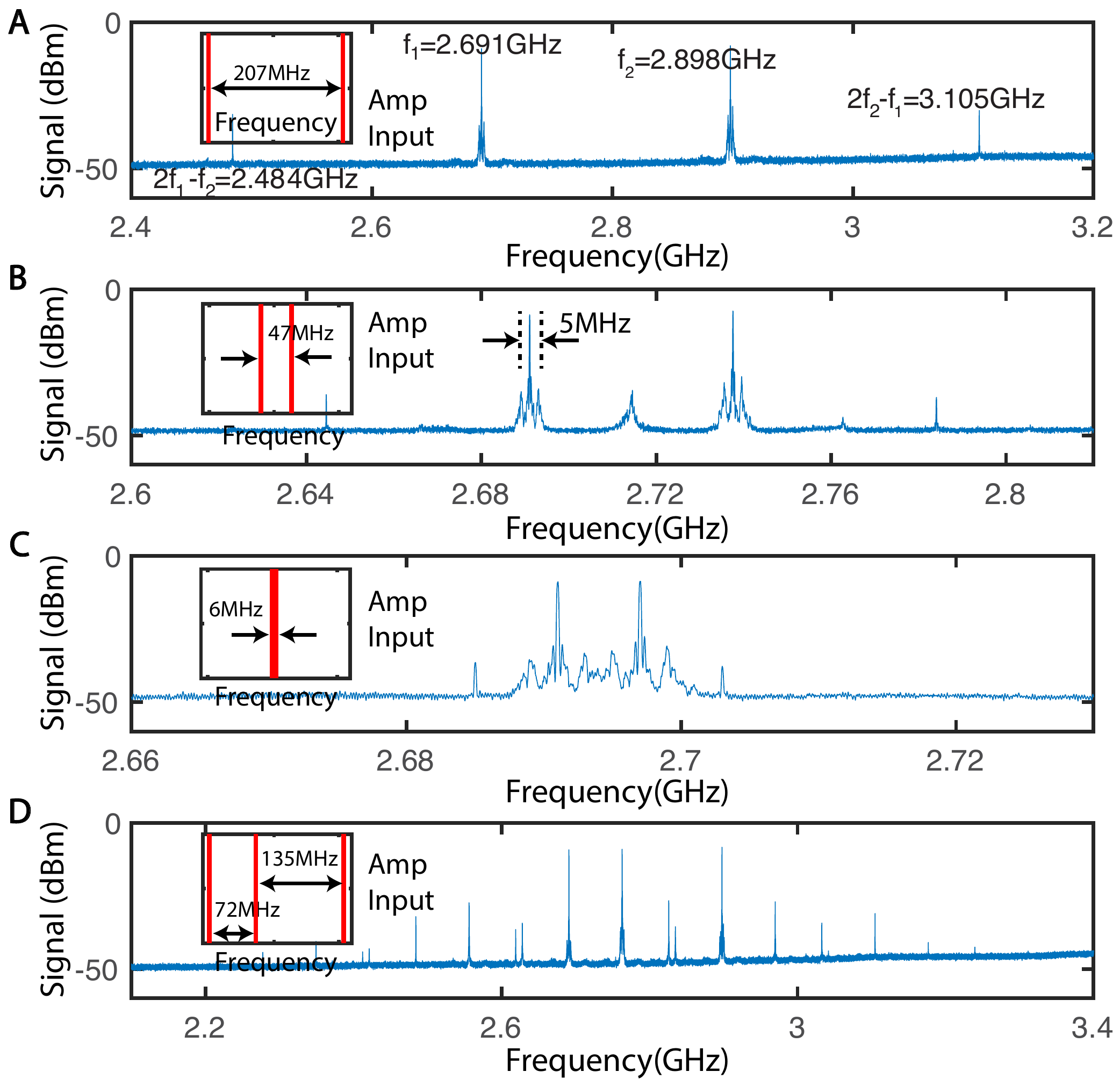}}
  \caption{
\textbf{Intermodulation distortion limits to cascaded sweeps}. Here we demonstrate amplifier nonlinearity effects on employing multiple cascaded sweeps on a single microwave amplifier (here Minicircuits ZHL-16W-43S+). We operate here in the linear region of the amplifier.  (A)-(C) Amplifier output spectrum with input of 2 frequency tones $f_1,f_2$ (insets) with decreasing separation. The third order IMD harmonics are clearly discernible, but are several orders of magnitude lower in power than the primary tones, and do not excite hyperpolarization. The nonlinearity however can cause severe distortion when the tones are $<$10MHz apart, or the amplifier is operating at saturation. (D) Amplifier output spectrum with 3 input frequency tones (insets). Several additional IMD harmonics are visible, although still at much reduced powers.
}
\zfl{IMD}
\end{figure}

\subsection{Intermodulation Distortion limits}
The multiplicative DNP gains of our microwave frequency comb technique will be limited by hyperfine mediated electron broadening, and also practical constraints such as non-linear distortions set by amplifier intermodulation distortion (IMD). Here we highlight the latter and discuss its effect on our hyperpolarization technique.

IMD appears in a wide range of RF and microwave systems, and particularly affects our experiments in the power amplifier stage \zfr{DNP-circuit}. When input a two-tone signal, an ideal linear amplifier would produce an output signal of the amplified two tones at exactly the same frequencies as the input signal. A realistic amplifier, however, will produce additional signal content at frequencies other than the two input tones. As one can tell from \zfr{IMD}, two fundamental tones, $f_1$ and $f_2$, injected into the amplifier mix to produce interfering signals with the most notable interference given by third order products, which are  $2f_1 -f_2$ and $2f_2 -f_1$. The power in these intermodulation products depends on how close the amplifier is to saturation. After saturation, the gain becomes nonlinear and enters a compression regime where the output power becomes independent of the input power. In this compression regime, intermodulation products and distorted signals arise from the mixing of fundamental signals which can adversely alter the signals of the amplified bandwidth. Since in our DNP mechanism the transfer efficiency falls rapidly with MW power, as long as one operates far below the amplifier compression point, these harmonics do not play a significant role in hyperpolarization process. However, when the two tones approach a frequency separation $<$10MHz (\zfr{IMD}C), the nonlinearity of the amplifier substantially distorts the output signal and deleteriously affects the DNP efficiency. In general, cascading more frequency sweeps lead to a larger number of spurious IMD harmonics, all of which take away MW power from the main frequency comb that drives the hyperpolarization process. This is evident for instance in \zfr{IMD}D, where we consider three frequency tones.

This technical obstacle can be overcome by employing multiple cascaded amplifiers, each amplifying a component of the frequency comb, which are then subsequently combined. This exploits the fact that power splitters have significantly lower nonlinearity, are not prone to IMD, and can yield a distortion-free combination of frequency tones. While the overall MW power increases linearly with the number of comb teeth, the ability to combine several low-power amplifiers to obtain multiplicative gains in DNP enhancements also has serious advantages in overall cost of the electronic infrastructure required.

\subsection{MW frequency combs at high field}
Let us now describe how the swept frequency combs can be constructed in the context of high field DNP with radicals. The availability of arbitrary waveform generators with bandwidths that reach into the microwave range, combined with solid-state millimeter wave mixers, now permits arbitrary pulse shaping up to Terahertz frequencies.  There are two approaches possible.  At frequencies below about 100GHz, it is possible to directly mix a high frequency carrier with a modulated microwave signal, filter it, and feed the resulting modulated millimeter waves into an amplifier.  Solid-state amplifiers with up to 1W of power are commercially available below 100GHz.  At even higher millimeter wave frequencies, an active multiplier chain (AMC-Virginia Diodes) is used to generate the millimeter wave signal, with a microwave input to the AMC in the 10-20GHz range.    These AMC’s are available with up to about 100mW of power at 270GHz, dropping to about 10mW at 500GHz and 1mW at 1THz.  By modulating the input to the AMC, it is possible to create arbitrarily modulated millimeter waves.  Note that in this case it is necessary to scale the desired modulation down by the multiplication factor of the AMC. 

The use of solid-sources at higher frequencies is typically constrained by the available power and the significant increase in cost with increased power output.    At liquid helium temperatures, the electron spin relaxation times become much longer, and it is possible to excite DNP at lower millimeter wave power.   At liquid nitrogen temperatures typically used for DNP-MAS experiments, the electron spin relaxation times are short, and high millimeter wave powers are needed to ensure good DNP.  Gyrotrons are used to generate sufficient millimeter wave power for DNP in this regime.  While gyrotrons have typically been narrow band (resonant devices), voltage-tunable gyrotrons that enable frequency modulation are being developed~\cite{Hoff15}.

\section{Data Analysis}
The DNP enhancement in our experiments is quantified by scaling the hyperpolarized $\Cs$ signal  to the corresponding thermal signal at 7T for each sample. The spectra are all phased, baseline corrected, and scaled to have an average noise of 1. This allows comparison between signals taken with a different number of averages. The areas of each peak area was calculated, and the ratio between them determines the enhancement factor. Zero-order phase correction is applied by multiplying the spectrum by a phase value that maximizes peak height.

A fitted absorptive Lorentzian curve identifies the peak in each spectra with the real portion of the data. Standard Lorentzian formulas were used to calculate the area under the fitted curve, and peak limits were designated such that the area between the limits encompassed 90 percent of the entire spectrum area. The portion outside of the peak limits was defined as noise. To flatten the baseline, a 12th order polynomial was fitted through these parts, and subtracted from the spectrum. For comparison between DNP and thermal spectra, the average noise of both spectra was scaled to 1 by dividing the noise section by its standard deviation. The area of each peak was then obtained through a Riemann sum across the peak limits, and the DNP enhancement factor given by the equation:

\beq
\varepsilon = \fr{\R{SNR}_\R{DNP}}{\R{SNR}_\R{Thermal}}\:\sq{\fr{N_{\R{DNP}}}{N_{\R{Thermal}}}}
\eeq

\end{document}